\documentclass[pra,twocolumn,superscriptaddress]{revtex4-2}

\usepackage{appendix}
\usepackage{orcidlink}

\usepackage{dsfont} 

\usepackage{mathtools}
\usepackage{siunitx}
\usepackage{upgreek}
\usepackage{bbold}
\usepackage[caption=false]{subfig}
\usepackage{graphicx}
\usepackage{hyperref}
\usepackage[capitalise]{cleveref}

\newcommand{\ii}[0]{\mathrm{i}}
\newcommand{\dd}[0]{\mathrm{d}}
\newcommand{\ee}[0]{\mathrm{e}}

\DeclareMathOperator{\Trace}{Tr}
\newcommand{\primesum}{\sideset{}{'}\sum}

\AtBeginDocument{\renewcommand{\vec}[1]{\mathbf{#1}}}
\AtBeginDocument{\renewcommand{\tensor}[1]{\underline{#1}}}

\begin{document}

\title{The Casimir-Polder interaction between atoms and hollow-core fibers}%

\author{B. Beverungen\orcidlink{0000-0002-6701-4269}}%
\affiliation{Humboldt-Universit\"at zu Berlin, Institut f\"ur Physik, 12489 Berlin, Germany}
\email[Corresponding author: ]{bettina.beverungen@physik.hu-berlin.de}
\author{D. Reiche\orcidlink{0000-0002-6788-9794}}%
\affiliation{Humboldt-Universit\"at zu Berlin, Institut f\"ur Physik, 12489 Berlin, Germany}
\author{K. Busch\orcidlink{0000-0003-0076-8522}}%
\affiliation{Humboldt-Universit\"at zu Berlin, Institut f\"ur Physik, 12489 Berlin, Germany}
\affiliation{Max-Born-Institut, 12489 Berlin, Germany}
\author{F. Intravaia\orcidlink{0000-0001-7993-4698}}%
\affiliation{Humboldt-Universit\"at zu Berlin, Institut f\"ur Physik, 12489 Berlin, Germany}
\date{\today}%

\begin{abstract}

The Casimir-Polder force acts on polarizable particles due to quantum fluctuations of the electromagnetic field that are modified by the presence of material bodies. We investigate the Casimir-Polder interaction for atoms near cylindrical fibers with hollow cores. 
This geometry represents one of the archetypal configurations encountered in numerous experimental setups designed to control and manipulate atoms in fundamental and quantum technological applications. 
Specifically, we analyze how the interplay of both geometrical and material-related length scales characterize the interaction, emphasizing the impact of the shell thickness. 
We develop a flexible and fast-converging numerical scheme for evaluating the interaction over a wide range of atom-cylinder separations at both zero and finite temperature. 
Furthermore, we provide a detailed analytical investigation of how various material properties modify the Casimir-Polder potential.
Finally, we analyze and discuss a number of limiting cases and compare numerical computations with corresponding analytical asymptotic expressions. 
In particular, in this geometry the Casimir-Polder potential is able to distinguish between an ohmic and non-ohmic description of conductors. 
One of the most significant outcomes of our work is that the shell thickness emerges as a useful parameter for controlling the interaction, opening avenues for both fundamental physics and applications in quantum technologies.
\end{abstract}

\maketitle

\section{Introduction}
\label{sec:introduction}

When the electromagnetic environment of an atom in vacuum is altered by bringing it into close proximity of a macroscopic object, a modification of the atomic energy levels occurs. 
Similar to the Lamb shift~\cite{Lamb47}, this phenomenon originates from the quantum fluctuations of the electromagnetic field. 
However and contrary to the Lamb shift, the broken translational invariance originating from the presence of the object leads to a position-dependence of the atomic level structure. 
This phenomenon, known as the Casimir-Polder effect~\cite{Casimir1948}, results in a quantum mechanical interaction that typically attracts the atom towards the object but also strongly depends on both the object's geometry and material composition.

The Casimir-Polder force can be characterized by a thermodynamic potential, a free energy, which typically decreases with increasing separation between the atom and the object, featuring different regimes with characteristic power-law dependencies with respect to the inverse separation. 
This phenomenon is relevant in all experimental setups designed to investigate or manipulate atoms positioned close to material structures.
Notable examples include atom chips~\cite{Reichel99,Fortagh07,Henkel09,Keil16} and atoms trapped near fibers~\cite{Vetsch10,Solano17a,Le-Kien18,Langbecker17,LeKien2022,Pennetta2025} or carbon nanotubes~\cite{Fermani07,Gierling11,Schneeweiss12}. 
In these cases, the Casimir-Polder force imposes constraints on the trapping distance~\cite{Folman02,Vetsch10} while simultaneously generating intriguing phenomena that influence the atomic dynamics~\cite{Hunger10,Hummer21}.
These systems are not only relevant for fundamental investigations~\cite{Lodahl17,Meng18,Hummer21,Lechner23} and also play a central role in quantum technologies.
Consequently, developing a comprehensive quantitative understanding is central for controlling the Casimir-Polder interaction and the advancement of future quantum devices.
\begin{figure}[t]
  \centering
  \includegraphics[width=0.48\textwidth]{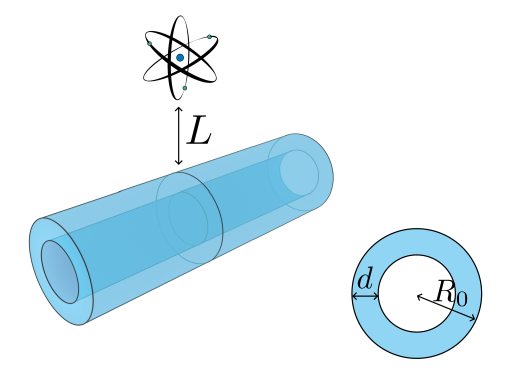}
  \caption{The system considered in this work. An atom in vacuum interacting with an infinitely extended cylindrical shell of radius $R_{0}$, thickness $d$ and comprised of different materials. The behavior of the Casimir-Polder free energy is described as a function of the atom-shell separation $L$.
   \label{fig:geometry-configuration} }
\end{figure}

The sensitivity of the Casimir-Polder force with regards to the properties of the atom's environment, in particular the geometry and material composition of nearby objects, is one of the interaction's most intriguing aspects. This is exemplified
when one or several geometrical dimensions are restricted, such as for slabs~\cite{Bostrom2000,Contreras-Reyes09,Dufour13,Buhmann2012,Buhmann2012a}, cylinders~\cite{Nabutovskii1979,Marvin1982,Barash1988,Barash1990} or spheres~\cite{Nabutovskii1979,Marvin1982,Buhmann2012,Buhmann2012a}. 
In all these cases, the interplay between geometry and material properties facilitates a far-reaching control of the interaction.
The case of an atom near a cylindrical structure (see Fig.~\ref{fig:geometry-configuration}) is of particular interest since it represents an archetypal configuration that relates to a number of the 
above-mentioned experimental setups~\cite{Note1}.
An infinitely extended cylinder can indeed represent optical fibers, nanowires, or other nanostructures (e.g.\ carbon nanotubes) present in the system.
Due to the intricate mathematics, the literature describing the Casimir-Polder interaction for this geometry often foregoes a numerical evaluation or makes various simplifications, such as the assumption of perfectly conducting cylinders (see, for instance, Refs.~\cite{Eberlein2007,Eberlein2009,Bezerra2011}). 
Employing idealized material properties like perfect conductivity have proven to be a rather successful approach for planar geometries (as exemplified by the original calculations by Casimir and Polder~\cite{Casimir1948a,Casimir1948a}). 
However, as we shall discuss below, for cylinders this approximation provides only a rough account of the interaction’s characteristics, significantly failing in certain circumstances. 
Early studies~\cite{Zeldovich1935,Nabutovskii1979,Marvin1982,Barash1989} relied on diverse analytical approaches with varying levels of detail.
More recent work employs numerical methods~\cite{Stourm20,LeKien2022}, for which identifying the system’s distinctive features and 
their physical interpretation can be more involved.

The physics of the system becomes even more interesting when additional geometrical features are introduced. 
Previous investigations have shown that fluctuation-induced interactions, such as the Casimir-Polder force, are very sensitive to systems comprised of layers of finite thickness~\cite{Bostrom00b,Pirozhenko08,Contreras-Reyes09}. 
When combined with a cylindrical structure this geometrical feature not only offers an additional knob for tailoring the interaction, but also allows for a more precise description of realistic setups. 
Indeed, in some of the previously considered experimental configurations, fibers and nanowires are coated with different materials with respect to their core~\cite{Schutz14} and more generally, layered cylindrical structures and, specifically, hollow-core cylinders are common in experiments~\cite{Schmidt13,Epple14,Bykov15}. 
From a general perspective, the interplay between the thickness of the cylindrical layers, the intrinsic and apparent dimensionality of the system (depending on the approximations), as well as the other length scales introduced by a specific material description can offer new pathways to modify the behavior of the interaction in unexpected ways.
Investigations of the Casimir-Polder force in cylindrical core-shell systems are relatively scarce in the literature~\cite{Blagov2005} and they are often framed in the context of simplifying schemes such as the so-called proximity force approximation~\cite{CasimirPhysics11}: Within this approach, which describes the surface as a collection of locally parallel surface elements, the impact of its curvature cannot be considered to its full extent. 
Similarly, fully numerical analyses are difficult to be found, largely because the number of quite distinct length scales that characterize the system presents significant challenges regarding convergence and required resources.

In this work, we aim to provide a comprehensive analytical and numerical investigation of the Casimir-Polder interaction between an atom and a cylindrical shell for a wide range of parameters. 
To elucidate the distinctive properties of the structure, we conduct an analysis of configurations involving several realistic material models such as dielectrics, metals, and non-dissipative conductors.
Following the most recent literature~\cite{Intravaia2011,Buhmann2012,Buhmann2012a}, our approach centers on the computation of the electromagnetic Green tensor for a cylindrical shell and we place particular emphasis on the characteristics of the material comprising the structure. In fact, different materials introduce different additional length scales that significantly impact the interaction.
A pivotal quantity in this analysis is the penetration depth of the electromagnetic field for a specific material. In this context, we primarily distinguish between conductors and dielectrics.
It is noteworthy that many of our results are independent of any specific model describing these materials and solely rely on their inherent physical properties.
In particular, we demonstrate that, in certain regions of distances, tunable through the shell thickness, the Casimir-Polder free energy behaves differently depending on the ohmic characteristics of the conductor.
This outcome can be of relevance in addressing a longstanding controversy surrounding the Casimir interaction between metallic plates~\cite{Klimchitskaya22c}.

Our manuscript is structured as follows.
In \cref{sec:casimir-polder-interaction-in-multilayered-systems} we first review the Green tensor based semi-analytical calculation of the Casimir-Polder interaction, with a particular focus on the geometry considered in this work. 
We provide compact expressions for all the key quantities characterizing the system.
In \cref{sec:numerical-evaluation}, we present a numerical approach that enables the quantitative analysis of the Casimir-Polder free energy for various regimes of the atom-cylinder separation.
Utilizing a recently developed Gaussian summation scheme, this approach facilitates the efficient and accurate computation of the Casimir-Polder interaction in systems characterized by cylindrical geometries across a wide range of length scales. 
In \cref{sec:zero-temperature-interaction} we provide asymptotic analyses which allow us to investigate the interplay of geometry and material properties for both dielectric and conducting shells. 
The asymptotic expressions not only provide a comprehensive description of the interaction’s behavior but also validate our numerical investigations.
While in \cref{sec:casimir-polder-interaction-in-multilayered-systems,sec:numerical-evaluation,sec:zero-temperature-interaction}
we develop the theory with a focus on the zero temperature case, we address finite-temperature corrections in \cref{sec:finite-temperature-interaction}.
The results are summarized and discussed in \cref{sec:conclusion}.

\section{Material and geometry influence on the Casimir-Polder interaction} 
\label{sec:casimir-polder-interaction-in-multilayered-systems}

We examine the interaction between a microscopic object (hereafter often referred to as an atom) and a macroscopic object consisting of a
material with spatially local but potentially frequency-dependent relative permittivity $\epsilon(\omega)$ surrounded by vacuum~\cite{Intravaia2011,Buhmann2012,Buhmann2012a,CasimirPhysics11}. 
Without the need to specify any particular geometry, to second order in the atom-field coupling strength the Casimir-Polder (free) energy for $T = \SI{0}{\kelvin}$ can be written as~\cite{Intravaia2011,Buhmann2012,Buhmann2012a}
\begin{equation}
\label{eq:cp-interaction-zero-temperature}
\mathcal{F}
=
-\hbar \int\limits_{0}^{\infty} \frac{\dd{\xi} }{2\pi} \,
  \Trace\left[\tensor{\alpha}(\ii\xi) \tensor{G}(\vec{r}_{a}, \vec{r}_{a}, \ii\xi)\right]~,
\end{equation}
where $\vec{r}_{a}$ is the position of the atom, $\hbar$ is the reduced Planck constant,  
and the trace ($\Trace$) runs over the three dimensions of space. 
The integrand is evaluated at imaginary frequencies and contains the product of the atom's polarizability tensor $\tensor{\alpha}(\omega)$ and the scattered part of the electric field Green tensor $\tensor{G}(\vec{r}, \vec{r'}, \omega)$~\cite{Intravaia2011,Buhmann2012,Buhmann2012a} (see \cref{app:green-tensor}). 
The latter encodes the information about the scattering properties of the macroscopic object in response to a point dipole source and is determined by both its geometry and material composition~\cite{Tai1994}. 
In \cref{eq:cp-interaction-zero-temperature}, the Green tensor is evaluated at coincidence, where both the source position $\vec{r}'$ and observation point $\vec{r}$ coincide with the position of the atom.
\Cref{eq:cp-interaction-zero-temperature} is a special case of the more general result for finite temperature $T \neq 0$, which can be expressed as a sum over the Matsubara frequencies $\xi_n = 2\pi n k_{B}T / \hbar$ ($k_{B}$ is the Boltzmann constant)
\begin{equation}
\label{eq:cp-interaction-finite-temperature} 
\mathcal{F}
=
- k_B T \primesum_{n=0}^{\infty} 
 \Trace\left[\tensor{\alpha}(\ii\xi_n) \tensor{G}(\vec{r}_{a}, \vec{r}_{a}, \ii\xi_n)\right]~,
\end{equation}
where the prime indicates that the $n=0$ summand is to be taken with half its weight. 
In the following, we initially focus our analysis on the $T = \SI{0}{\kelvin}$ case and investigate the modifications introduced by a finite temperature in detail in \cref{sec:finite-temperature-interaction}.

For simplicity, we assume hereafter that the atom's polarizability is isotropic, i.e. $\tensor{\alpha}(\omega) = \alpha(\omega) \mathds{1}$. 
The atom's internal dynamics are further characterized in terms of the polarizability's static value $\alpha_{0}\equiv\alpha(0)$ and a typical frequency $\omega_{a}$ (commonly in the optical range) which corresponds to the dominant atomic transition. 
As a consequence, along the imaginary frequency axis $\alpha(\ii \xi)$ becomes negligible for $\xi \gg \omega_{a}$.
Although most of our analysis does not depend on a specific form of the polarizability, in our numerical evaluations we shall consider the following expression
\begin{equation}
\label{eq:polarizability-model} 
\alpha(\ii\xi) = \alpha_{0} \frac{\omega_a^2}{\omega_a^2 +\xi^2}~.
\end{equation}
The polarizability introduces $\lambdabar_{a}=c/\omega_{a}$ as an additional relevant length scale for our system, i.e. the (reduced) wavelength corresponding to the dominant atomic transition frequency.
In relation to $\mathbf{r}_{a}$ and other material-specific characteristics, it can be used to distinguish between the nonretarded and the retarded region, determined by those atom-object separations where the effects of retardation induced by the finiteness of the speed of light can or cannot be neglected.

\Cref{eq:cp-interaction-zero-temperature,eq:cp-interaction-finite-temperature} are still rather general since an explicit expression for the Green tensor of the macroscopic object has not yet been specified. 
Before approaching the definition of this quantity for a cylindrical shell, it is helpful to first turn our attention on the characterization of the material properties, formulating their impact on our system in a way which is suitable for our analysis.  

\subsection{Electromagnetic field penetration depth in dielectrics and conductors}
\label{subsec:em-field-penetration-depth}

In the following analysis, we find it convenient to introduce what we call the \emph{relative} penetration depth
\begin{equation}
\label{eq:relative-penetration-depth}
\Delta\equiv\Delta(\ii \xi)=\frac{1}{\sqrt{\epsilon(\ii \xi)-1}}~,
\end{equation}
which is related to the common concept of electromagnetic penetration depth~\cite{Jackson75}. 
In fact, the dimensionless quantity in \cref{eq:relative-penetration-depth} is effectively the ratio between the penetration depth of the electromagnetic field in a material and the wavelength of the corresponding radiation, all evaluated at imaginary frequencies.  
This quantity is central to our analysis, as it serves as the unifying element to develop physical explanations for the potentially quite different behavior of the Casimir-Polder interaction for dielectrics and several types of conductors.
In general, since in the time domain the electromagnetic response function of a material is real and causal, along the positive imaginary frequency axis the relative permittivity function is real, larger than one, and monotonically decreasing~\cite{Jackson75,Landau80}. Consequently, $\Delta(\ii \xi)$ is a real, positive and monotonically increasing function for $ \xi\ge0$. 
Physically, given the high frequency transparency of any material and the fact that the optical response is a continuous function of time, at high imaginary frequency, $\epsilon(\ii \xi)-1$ vanishes as $\xi^{-2}$~\cite{Jackson75}, indicating that $\Delta(\ii \xi)\propto \xi$ for $\xi \to \infty$. 
Phenomenologically, however, permittivity models exist that well describe the material properties within a defined frequency range of frequencies but deviate from this constraint. For instance, $\epsilon(\ii \xi)-1$ can tend to a positive constant for $\xi\to \infty$, and $\Delta$ plateaus at large $\xi$ instead of diverging. Although it might be relevant in some other circumstances, this deviation does not impact the following considerations.
For our analysis, the low frequency regime is more pertinent, where $\Delta(\ii \xi)$ exhibits distinct characteristics depending on the material being examined.
While the electromagnetic field can deeply penetrate dielectrics, for conductors the opposite is true.
For example, a perfect electric conductor has $\Delta=0$.
Whereas $\epsilon(\ii \xi)$ tends to a constant in the limit $\xi \to 0$ for dielectrics, the relative permittivity of conductors features a divergence at low frequencies~\cite{Jackson75,Landau80}.
Physically, we can further distinguish between ohmic and non-ohmic conductors by virtue of their conductivity.
While the conductivity of the former typically approaches a constant at low frequencies, the zero-frequency conductivity behavior of non-ohmic conductors might vanish or diverge. 
Although various non-ohmic materials exist~\cite{Ma20}, and the formalism described below can be generalized to include them, we focus here on superconductor-like models, for which the low-frequency conductivity diverges, and the deviation from ohmic behavior is directly linked to the absence of dissipation.

As a consequence of the preceding discussion, the relative penetration depth $\Delta$, defined in \cref{eq:relative-penetration-depth}, exhibits a distinct behavior for small values of $\xi$. Specifically, depending on the material being considered, there exists a value $\xi_{0}$ for which we can write  
\begin{align}
\label{eq:relative-penetration-depth-materials-zero-frequency}
\Delta(\ii \xi)
\stackrel{\xi \ll \xi_{0}}{\sim}
\begin{dcases}
\Delta_{0}& \text{dielectrics}
\\
\sqrt{\frac{\lambdabar_{D}}{c}\xi}& \text{ohmic conductors}
\\
\frac{\lambdabar_{p}}{c}\xi& \text{lossless conductors}
\end{dcases}
~.
\end{align}
Here, $\lambdabar_{D}$ and $\lambdabar_{p}$ are length scales associated with the diffusion length of the electromagnetic field within the ohmic material, and the penetration depth in the non-dissipative conductor, respectively. 
Large values of $\lambdabar_{D}$ are typical for conductors with high resistivity, corresponding to a more pronounced ohmic behavior.
In contrast, small values of $\lambdabar_{p}$ characterize a non-dissipative conductor which behaves more like a perfect electric conductor. 
Commonly, $\lambdabar_{p}$ ranges between a few tens or a few hundreds of nanometers. 
The value of $\lambdabar_{D}$ can vary between a few Angstrom for good conductors to a few tens of nanometers for poor conductors. 
The value of $\Delta_{0}$ scales as the inverse of the static dielectric constant, with typical values of the order of one for common dielectrics. 

Although, as mentioned, most of our theoretical analyses are independent of any specific expression for the permittivity, the numerical evaluations in our work employ a single Lorentz oscillator model for dielectrics, while the Drude model and its dissipationless limit are utilized for ohmic and non-ohmic conductors, respectively. 
This corresponds to the following explicit material descriptions
\begin{equation}
\label{eq:permittivity-models}
\epsilon(\ii \xi) 
=
\begin{dcases}
\epsilon_{\infty} + \frac{\omega_p^2}{\xi^2 + \gamma\xi + \omega_0^2}& \text{dielectrics}
\\
\epsilon_{\infty} + \frac{\omega_p^2}{\xi^2 + \gamma\xi}& \text{ohmic conductors}
\\
\epsilon_{\infty} + \frac{\omega_p^2}{\xi^2}& \text{lossless conductors}
\end{dcases}~.
\end{equation}
In these expressions, $\epsilon_{\infty}\ge 1$ represents the standard background permittivity, a constant that incorporates the contribution to the dielectric function of processes within the materials occurring at frequencies significantly exceeding the optical range~\cite{Jackson75}. 
Along the positive imaginary frequency axis, where the permittitivity is monotonically decreasing starting from its maximum value at $\xi=0$, the constant offset introduced by an $\epsilon_{\infty} > 1$ results in the above-mentioned behavior of $\epsilon(\ii \xi)-1$ approaching a positive constant as $\xi \to \infty$. 
For our purposes, this becomes negligible due to the polarizability effectively restricting the relevant frequencies to $0<\xi\lesssim\omega_a$. 
However, the shift in permittivity within this range, and thus the value of $\epsilon_{\infty}$ itself can be quantitatively relevant to our analysis. 
The constant $\omega_{0}$ is a characteristic frequency of the dielectric material, while $\omega_p$ denotes the plasma frequency and $\gamma$ the phenomenological damping rate for both kind of materials~\cite{Jackson75}. 
We would like to note, that within these models, $\xi_{0}=\omega_{0}$ corresponds
to the dielectric, $\xi_{0}=\gamma$ yields the conductor and $\xi_{0}= \omega_{p}/\sqrt{\epsilon_{\infty}-1}$ represents the non-dissipative conductor. Further, we have that $\lambdabar_{p}=c/\omega_{p}$, which is the reduced plasma wavelength, as well as $\lambdabar_{D}=c\gamma/\omega_{p}^{2}$ and $\Delta_{0}=1/\sqrt{\epsilon_{\infty} - 1 + \omega_{p}^2/\omega_{0}^2}$.
It is also opportune to mention that, mathematically, the description of the non-dissipative conductor is obtained from that of the dissipative metal by setting $\gamma = 0$.
For $\epsilon_{\infty} = 1$, our expression reduces to the London model~\cite{London35,Bimonte10}. 
As one of the earliest descriptions of superconductivity, this model captures the non-dissipative character of an ideal metal, though it fails to account for other phenomena such as the Meissner effect. 
In the context of our study, we are concerned solely with the non-dissipative characteristics of the material.
In the literature, this non-dissipative permittivity is frequently referred to as the plasma model~\cite{CasimirPhysics11}. 
It can also be thought of as a high frequency ($\xi \gg \gamma$) approximation of the Drude model, and its use is often sufficient in practical calculations, although there are notable exceptions where results differ significantly (see \cref{subsubsec:thin-wire-T-zero-conductors,sec:finite-temperature-interaction}).

\subsection{Scattering properties of a hollow-core fiber}
\label{subsec:em-field-scattering-cylinder}

Geometry-related effects enter the Casimir-Polder interaction in \cref{eq:cp-interaction-zero-temperature}
via the Green tensor, which contains the information about the scattering properties of the macroscopic object. 
In this work, we specifically focus on the experimentally relevant geometries of hollow-core fibers and nanowires. 
If the structure's distance to the atom is much smaller than its length, 
it can be idealized as an infinitely extended cylindrical shell (cf. \cref{fig:geometry-configuration}).
This is one of the few highly symmetric geometries for which a semi-analytic formula of the corresponding Green tensor can be obtained~\cite{Tai1994, Tomas1995, Li1995, Li2000, Li1994}. 
The underlying strategy to obtain such an expression involves solving the vector wave equation with a point-like dipole source, and expanding the solution in vector wave functions which are adapted to the specific geometry (see \cref{app:green-tensor}), e.g. plane waves for planar systems and cylindrical vector waves for cylindrical ones. 
With this choice, the boundary conditions the electromagnetic field is required to satisfy at the interface between two different media~\cite{Jackson75} can be readily implemented~\cite{Tai1994, Tomas1995}. 
This leads to the definition of the scattering coefficients for a single isolated interface and eventually for the entire multilayer system~\cite{Tai1994}. 
These coefficients, which enter into the expression for the Green tensor, encode how light of a given polarization is scattered at an interface -- they are equivalent to the well-known Fresnel reflection coefficients for a planar, semi-infinite medium, or the Mie coefficients in the case of cylinders and spheres~\cite{Mie08,Tzarouchis18,Bohren08}.

For the cylindrical geometry which we consider in our work, we choose its axis of symmetry to coincide with the $z$-axis. 
We work in a cylindrical coordinate system with unit vectors $\vec{\hat{e}}_{i}$ where $i \in \{\rho, \phi, z\}$. 
Then, the structure's material composition is rotationally symmetric around the $z$-axis, while we allow for a multilayer composition along the radial direction, i.e.\ along $\vec{\hat{e}}_{\rho}$. 
To specify the hollow-core cylinder geometry illustrated in \cref{fig:geometry-configuration}, we denote the radius of the outermost surface as $R_{0}$ and assume that the structure is embedded in vacuum.
The internal radius $R_{i}$ fulfills $R_{i} < R_{0}$, which also allows us to define the thickness of the shell $d = R_{0} - R_{i}$. 
For reasons of symmetry, the position of the atom can be given in terms of the radial coordinate only, i.e. $\vec{r}_{a} = r_{a} \vec{\hat{e}}_{\rho}$, where we have assumed $z=0$ without loss of generality.  We write $r_{a} = R_{0} + L= (1+s) L$, where $L$ is the radial distance to the cylinder's surface and we have defined $s = R_{0} / L$. 
This last parameter is central to our evaluation and effectively accounts for the impact of the cylinder's curvature in all of the subsequent considerations. 
In the limit $s \gg 1$, all expressions must recover those of a planar structure. 
Since a hollow-core cylinder appears in this limit as a slab of finite thickness $d$, we designate this region of distances as the \emph{slab limit}. 
Specifically, for a perfectly conducting material, the thickness of the slab is irrelevant and we must recover the result of Casimir and Polder~\cite{Casimir1948,Intravaia2011,Buhmann2012,Buhmann2012a}. 
In the nonretarded ($L\ll\lambdabar_{a}$) and the retarded ($L\gg\lambdabar_{a}$) regime, the slab limit explicitly reads as
\begin{gather}
\label{eq:CasimirPolder48}
\mathcal{F}\sim
\begin{dcases}
-\frac{\hbar }{16 \pi^{2} L^{3}}\int\limits_{0}^{\infty}\dd{\xi} \frac{\alpha(\ii\xi)}{\epsilon_{0}}, & L\ll\lambdabar_{a}
\\
\\
- \hbar c\frac{3\alpha_{0}}{32 \pi^{2}\epsilon_{0} L^{4}}\equiv \mathcal{F}_{0},& L\gg\lambdabar_{a}
~.
\end{dcases}
\end{gather}
Conversely, for $s\ll 1$ we are in what we shall hereafter call the \emph{thin-wire limit}, since for these distances the hollow-core cylinder effectively appears as an infinite one-dimensional solid wire. 
The latter limit corresponds to a region of distances $L$ where the impact of the structure's geometrical confinement along the transverse direction is most prominent and this motivates the major part of the analytical investigations in this article (see \cref{subsec:thin-wire-T-zero,sec:finite-temperature-interaction}).

While we defer the explicit expression of the Green tensor to \cref{app:green-tensor}, we shall briefly summarize its main components insofar as they are relevant for the subsequent analysis.
The Green tensor can be conveniently written as an integral over $q$, i.e. the modulus of the wave vector component along $\vec{\hat{e}}_{z}$. 
It further contains an infinite series over the nonnegative integers indexed by $m$, which reflects the periodicity of the solution in the azimuthal direction.
Owing to the cylindrical symmetry of the problem, the expression involves the modified Bessel functions of the first and second kind of order $m$, $I_m$ and $K_m$, as well as their corresponding logarithmic derivatives, which we shall denote by $\mathbb{i}_m=I'_m/I_m$ and $\mathbb{k}_m=K'_m/K_m$~\cite{Abramowitz71}. 
In general, these functions depend on the radius $R_{0}$ and on $\kappa = \sqrt{q^2 + \xi^2/c^{2}}$, which corresponds to the propagation constant in vacuum related to the component of the wave vector orthogonal to the surface, evaluated at imaginary frequencies.

In accordance with our previous discussion, the cylinder's scattering coefficients appearing in the Green tensor effectively describe how the electromagnetic radiation is scattered from the outermost interface, and we denote them here by $r_{m}^{ij}\equiv r_{m}^{ji}$, ${i,j\in\{\mathrm{N},\mathrm{M}\}}$, and we provide the corresponding expressions in Eqs.~\eqref{eq:scattering-coefficients-hollow-cylinder}~\cite{Note2}.
%\footnote{Our notation slightly differs from the literature~\cite{Tai1994}, but is more suitable for our purposes.}
For multilayer cylindrical structures, the scattering coefficients are functions of the radii corresponding to the position of the interface layers, as well as the propagation wave vectors within each layer.
For example, in our geometry we have only one layer and, if the corresponding material is local and isotropic, in analogy to the vacuum case we can express the propagation constant at imaginary frequencies as $\kappa_{\epsilon} = \sqrt{q^2 + \epsilon(\ii \xi)\xi^2/c^{2}}$.

In the case of hollow-core cylinders, it is still practical to determine an explicit expression for the scattering coefficients (see for example, Ref.~\cite{Rekdal04} and more generally \cite{Tai1994}). However, differently from previous work on cylindrical structures, we reformulate these expressions to compactly write them as follows
\begingroup
\allowdisplaybreaks
\begin{subequations}
\label{eq:scattering-coefficients-hollow-cylinder}
\begin{spreadlines}{1em}
\begin{align}
r_{m}^{\mathrm{NN}}
&=\frac{\frac{\left(1+\frac{1}{\Delta^2}\right)\Phi_{m}^{(\epsilon)} -\frac{\eta_{\epsilon}}{\eta} \frac{\mathbb{i}_m(\eta s)}{\mathbb{i}_m(\eta_{\epsilon} s)}}{\Pi_{m}^{21}}
+
 \frac{ \left(\frac{\zeta\sqrt{1-\zeta^{2}} }{\Delta^{2}}\frac{\frac{m}{\eta_{\epsilon} s}}{\mathbb{i}_m(\eta_{\epsilon} s)} \right)^{2}
}{\Phi_{m}^{(1)}-\frac{\eta_{\epsilon}}{\eta}\frac{\mathbb{k}_m(\eta s)}{\mathbb{i}_m(\eta_{\epsilon} s)} }
}
{\frac{ \left(1+\frac{1}{\Delta^2}\right)\Phi_{m}^{(\epsilon)} -\frac{\eta_{\epsilon}}{\eta} \frac{\mathbb{k}_m(\eta s)}{\mathbb{i}_m(\eta_{\epsilon} s)}}{\Pi_{m}^{22}}
+
\frac{\left(\frac{\zeta\sqrt{1-\zeta^{2}} }{\Delta^{2}}\frac{\frac{m}{\eta_{\epsilon} s}}{\mathbb{i}_m(\eta_{\epsilon} s)}  \right)^{2}
}{\Phi_{m}^{(1)}-\frac{\eta_{\epsilon}}{\eta}\frac{\mathbb{k}_m(\eta s)}{\mathbb{i}_m(\eta_{\epsilon} s)}}
}
\\
r_{m}^{\mathrm{MM}}
&=\frac{
\frac{\Phi_{m}^{(1)}-\frac{\eta_{\epsilon}}{\eta} \frac{\mathbb{i}_m(\eta s)}{\mathbb{i}_m(\eta_{\epsilon} s) }}{\Pi_{m}^{12}}+  
\frac{\left( \frac{\zeta\sqrt{1-\zeta^{2}} }{\Delta^{2}}\frac{\frac{m}{\eta_{\epsilon} s}}{\mathbb{i}_m(\eta_{\epsilon} s)} \right)^{2}}
{\left(1+\frac{1}{\Delta^2}\right)\Phi_{m}^{(\epsilon)} -\frac{\eta_{\epsilon}}{\eta} \frac{\mathbb{k}_m(\eta s)}{\mathbb{i}_m(\eta_{\epsilon} s)}}}
{
\frac{\Phi_{m}^{(1)}-\frac{\eta_{\epsilon}}{\eta}\frac{\mathbb{k}_m(\eta s)}{\mathbb{i}_m(\eta_{\epsilon} s)}}{\Pi_{m}^{22}} +  
\frac{\left( \frac{\zeta\sqrt{1-\zeta^{2}} }{\Delta^{2}}\frac{\frac{m}{\eta_{\epsilon} s}}{\mathbb{i}_m(\eta_{\epsilon} s)} \right)^{2}}
{ \left(1+\frac{1}{\Delta^2}\right)\Phi_{m}^{(\epsilon)}-\frac{\eta_{\epsilon}}{\eta} \frac{\mathbb{k}_m(\eta s)}{\mathbb{i}_m(\eta_{\epsilon} s) }}}
\\
r_{m}^{\mathrm{MN}}
&=-\frac{\frac{1}{\Pi_{m}^{0}} 
\frac{\left(\frac{\zeta\sqrt{1-\zeta^{2}} }{\Delta^{2}} \frac{\frac{m}{\eta_{\epsilon} s}}{\mathbb{i}_m(\eta_{\epsilon} s)} \right)\left[
\frac{\eta_{\epsilon}}{\eta}\frac{\mathbb{i}_{m}(\eta s)}{\mathbb{i}_m(\eta_{\epsilon} s) }-
\frac{\eta_{\epsilon}}{\eta}\frac{\mathbb{k}_{m}(\eta s)}{\mathbb{i}_m(\eta_{\epsilon} s) }\right]}
{
 \left[\left(1+\frac{1}{\Delta^2}\right)\Phi_{m}^{(\epsilon)} -\frac{\eta_{\epsilon}}{\eta} \frac{\mathbb{k}_m(\eta s)}{\mathbb{i}_m(\eta_{\epsilon} s) }\right]
\left[\Phi_{m}^{(1)}-\frac{\eta_{\epsilon}}{\eta}\frac{\mathbb{k}_m(\eta s)}{\mathbb{i}_m(\eta_{\epsilon} s)} \right]}
}{
\frac{1}{\Pi_{m}^{22}}
+   
\frac{\left(\frac{\zeta\sqrt{1-\zeta^{2}} }{\Delta^{2}}
\frac{\frac{m}{\eta_{\epsilon} s}}{\mathbb{i}_m(\eta_{\epsilon} s)}
 \right)^2}
 { \left[\left(1+\frac{1}{\Delta^2}\right)\Phi_{m}^{(\epsilon)} -\frac{\eta_{\epsilon}}{\eta} \frac{\mathbb{k}_m(\eta s)}{\mathbb{i}_m(\eta_{\epsilon} s) }\right]
\left[\Phi_{m}^{(1)}-\frac{\eta_{\epsilon}}{\eta}\frac{\mathbb{k}_m(\eta s)}{\mathbb{i}_m(\eta_{\epsilon} s)} \right]}}~,
\end{align}
\end{spreadlines}
\end{subequations}
\endgroup
where, for simplicity, we dropped the arguments of all the involved functions.
In Eqs.~\eqref{eq:scattering-coefficients-hollow-cylinder}, we have introduced the dimensionless variables
\begin{equation}
\label{eq:dimensionless-variables-zero-temperature}
\eta = \kappa L, \quad \eta_{\epsilon} = \kappa_{\epsilon} L \quad \text{and} \quad \frac{\xi}{c}L =\zeta \eta
~.
\end{equation} 
The functions $\Phi_{m}^{(x)}$ and $\Pi_{m}^{ij}$ depend on $R_{i}$ and, for brevity, we report their expressions in \cref{app:green-tensor}.
In the limit $R_{i}\to 0$, they tend to one, thus prominently marking the difference between a cylindrical shell and a full cylinder.
We note that, differently from planar structures, a cylindrical vector wave with either of the two polarizations $i = \mathrm{N}, \mathrm{M}$ generally does not conserve its polarization state during the scattering process, giving rise to a cross-polarization scattering coefficient $r_{m}^{\mathrm{MN}}$~\cite{Tai1994}. 
It is worth to remark that Eqs.~\eqref{eq:scattering-coefficients-hollow-cylinder} can be readily simplified when $m=0$ and we notice that then $r_{0}^{\mathrm{MN}}$ identically vanishes, pointing to a total decoupling of the polarization at this order.

Using the above results and definitions, we can now write the Casimir-Polder free energy for a cylindrical structure at $T=\SI{0}{\kelvin}$ as
\begin{widetext}
\begin{align}
\begin{split}
\mathcal{F}
=
- 
\frac{\hbar c}{2 \pi^3} \frac{1}{L^4}
\int\limits_{0}^{\infty} \dd{\eta}
\int\limits_0^{1} \dd{\zeta}  
 \frac{ \alpha\left(\ii\zeta\frac{c}{R_{0}}\eta s\right)}{\epsilon_{0}}&
 \primesum_{m=0}^{\infty} \;
  \frac{\mathcal{K}_{m}(\eta,s)}{\sqrt{1-\zeta^2}} 
  \Bigg\{~
     r_{m}^{\mathrm{NN}}
  \left( 
     1 +
     (1-\zeta^{2}) \biggl[ \frac{m^2}{\eta^{2} [1+s]^2}
      +
     \mathbb{k}^{2}_m (\eta [1+s])
    \biggr]\right)
\\
  {}-{} &r_{m}^{\mathrm{MM}} \zeta^{2}\left[ 
     \frac{m^2}{\eta^{2} [1+s]^2} 
     +
     \mathbb{k}^{2}_m(\eta [1+s])
 \right]
  + r_{m}^{\mathrm{MN}}4\zeta\sqrt{1-\zeta^{2}}
  \frac{ m }{\eta [1+s]} \mathbb{k}_m (\eta [1+s]) 
  \Bigg\}~.
\end{split}
\label{eq:cp-interaction-dimensionless-text}
\end{align}
\end{widetext}
Here, the integration runs over the dimensionless variables introduced in \cref{eq:dimensionless-variables-zero-temperature}.
As above, the prime in the sum indicates that the term with $m=0$ must be weighted with a factor of $1/2$. 
The free energy ${\cal F}$ is defined in terms of the kernel function
\begin{equation}
 \mathcal{K}_{m}(\eta,s)=\eta^{3}\frac{I_m(\eta s )}{K_m(\eta s)} K_m^2(\eta [1+s])
~,
\label{eq:cp-kernel-function}
\end{equation}
which proves useful in the subsequent analyses.
It is opportune to underline that formally the expression in \cref{eq:cp-interaction-dimensionless-text} holds in general for an atom above any multilayered cylinder embedded in vacuum, assuming that the corresponding scattering coefficients $r_{m}^{ij}$ are known or can be determined numerically.

\section{Numerical evaluation}
\label{sec:numerical-evaluation}

To comprehensively investigate the physics at play in the system and to validate the analytical results presented below, we have implemented an accurate and efficient scheme for the numerical evaluation of the Casimir-Polder interaction in cylindrical geometries.
In the following, we briefly summarize the main challenges arising in the computation of \cref{eq:cp-interaction-finite-temperature,eq:cp-interaction-zero-temperature}, and give an overview of our approach. 
To streamline the discussion, we again focus on the zero-temperature case described by \cref{eq:cp-interaction-dimensionless-text} and defer all additional details as well as the treatment of finite-temperature effects to \cref{app:numerical-evaluation-details}.
 
An evaluation of the Casimir-Polder interaction between an atom in proximity to a cylindrical structure represents a computational challenge~\cite{Eberlein2007}. 
Essentially, this is rooted in both its mathematical complexity and the very distinct physical length scales  ($\lambdabar_{a}$, $R_{0}$, $d$, $L$, $\lambdabar_{D}$, $\lambdabar_{p}$, etc.) inherent to the physics underlying the problem.
The physically accessible values of these scales and their interrelations yield regimes with qualitatively rather different behavior of the interaction. 
As a result, a general purpose numerical evaluation scheme can become quite ineffective, especially in those regions where the relevant length scales differ by several orders of magnitude. 
To calculate the interaction more efficiently, it is thus beneficial to explicitly take into account the asymptotic behavior of the corresponding expressions.  
For cylindrical geometries, one especially relevant case is the limit of small aspect ratios, corresponding to $s\gg 1$. 
In this regime, the atom's distance to the surface is much smaller than the cylinder’s
radius, making the cylinder look almost flat to the atom. 
As a consequence, many terms in the sum in \cref{eq:cp-interaction-dimensionless-text} have to be considered in order to achieve convergence. 
In terms of the previously discussed choice of functions for the implementation of the boundary conditions, this can be understood as essentially having to represent a nearly planar wavefront in a cylindrical basis and obviously this is not the natural coordinate system for this task. 
Similarly, the calculation of finite-temperature effects according to \cref{eq:cp-interaction-finite-temperature} is known to exhibit slow convergence in the limit of low temperatures, requiring many terms of the Matsubara sum for obtaining reasonable accuracy. 
In this work, we approach the evaluation of the combined integration and summation in \cref{eq:cp-interaction-dimensionless-text} with a multivariate Gaussian quadrature scheme, using a suitably chosen fixed-point quadrature rule in each dimension (see e.g. \cite{Abramowitz71,Press2007}). 
In particular, we find that the aforementioned issues of slow convergence of the infinite series 
can be addressed effectively with a recently developed Gaussian summation scheme based on modified discrete Laguerre polynomials~\cite{Xu2021}.
The latter can be seen as a generalization of standard Gaussian quadrature schemes to infinite series, effectively corresponding to the evaluation of an integral with respect to a discrete measure~\cite{Engblom2006,Monien2010}. 

We now proceed to discuss the main points in reformulating \cref{eq:cp-interaction-dimensionless-text} to make it suitable for numerical evaluation with Gaussian quadrature and refer to \cref{app:numerical-evaluation-details} for further details. 
As a first step, we focus on the integrals over the dimensionless variables $\eta$ and $\zeta$ defined in \cref{eq:dimensionless-variables-zero-temperature}. 
The integral over $\zeta$ can be brought to the standard interval $[-1,1]$ by performing the change of variable  
$\tilde{\zeta} = 2\zeta-1$. 
The integrand possesses an integrable square root singularity at $\zeta=1$, which can be taken into account explicitly via a Gauss-Jacobi quadrature rule in the transformed variable $\tilde{\zeta}$.  
In order to efficiently compute the $\eta$ integral over the semi-infinite interval $[0,\infty)$, it is useful to analyze the asymptotic behavior of the integrand for large $\eta$. 
This behavior is determined by the kernel function in \cref{eq:cp-kernel-function}, yielding an exponential decay when $\eta$ is sufficiently large. 
The latter is a direct consequence of the modified Bessel functions' asymptotic exponential behavior for large values of their arguments~\cite{Abramowitz71}. 
For the numerical evaluation, this suggests to use the exponentially scaled modified Bessel functions $\tilde{I}_m(x)=I_m(x)e^{-x}$ and $\tilde{K}_m(x)=K_m(x)e^{x}$, available from standard numerical libraries~\cite{Amos1986}. 
In this way, the kernel function in \cref{eq:cp-kernel-function} can be replaced by its exponentially scaled counterpart $\tilde{\mathcal{K}}_{m}(\eta, s)$ according to
\begin{equation}
\mathcal{K}_{m}(\eta, s) 
= 
\tilde{\mathcal{K}}_{m}(\eta, s) e^{-2\eta}
~,
\label{eq:kernel-function-exponentially-scaled}
\end{equation}
thereby making explicit the exponential decay in $\eta$.  
As an additional benefit, the exponentially scaled modified Bessel functions of the first and second kind are less prone to floating point over- or underflow for large arguments, respectively. 
In this form, the integrand becomes amenable to integration over $\tilde{\eta}=2\eta$ with a Gauss-Laguerre quadrature rule. 
\begin{figure*}
\subfloat[]{
  \includegraphics[width=0.99\columnwidth]{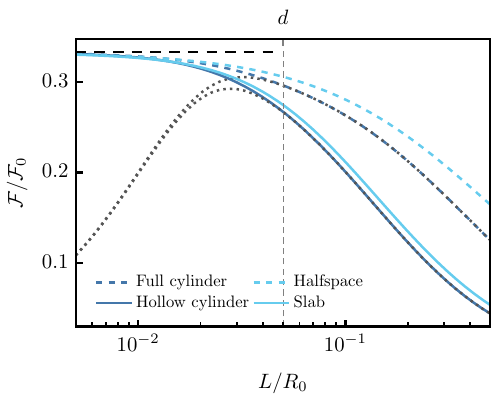}
}
\hspace{0.02\columnwidth}
\subfloat[]{
  \includegraphics[width=0.99\columnwidth]{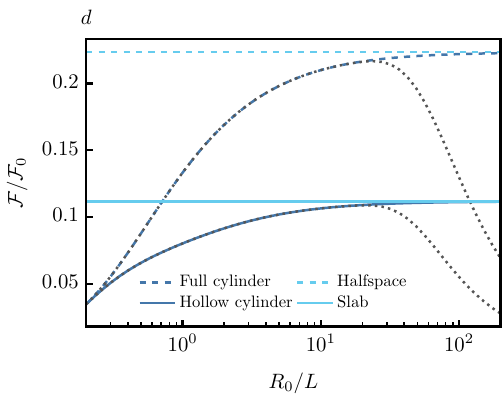}
}
\caption{Comparison of direct summation (dotted dark gray lines) and MDL quadrature (dark blue lines, see \cref{sec:numerical-evaluation}) for the evaluation of the zero temperature Casimir-Polder potential of an atom interacting with cylindrical geometries according to \eqref{eq:cp-interaction-dimensionless-text}. 
Specifically, the system consists of a Rubidium atom ($\omega_{a} = \SI{1.6}{\eV}$, $\lambdabar_{a}\approx \SI{120}{\nano\meter}$~\cite{Steck2015}) interacting with a dielectric cylindrical structure (similar results are achieved for conductors). The frequency dependent behavior of the relative permittivity is described using the first model of \cref{eq:permittivity-models} with values for silica ($\epsilon_{\infty}=1.49$, $\omega_p=\SI{7.8}{eV}$, $\omega_0=\SI{10}{eV}$ and $\gamma=\SI{0.9}{\eV}$~\cite{Pflug2022}). The potential is normalized by the nonretarded limit of the Casimir-Polder energy between an atom and a perfectly reflecting surface (see \cref{eq:CasimirPolder48}).
The interaction for the cylindrical geometries is depicted as a) a function of $L$ for a fixed external radius $R_{0}=2 \lambdabar_{a}$ and b) as a function of $R_{0}$ at a fixed distance $L=0.5 \lambdabar_{a}$. 
Two values of internal radii are used: $R_{i}=0$, i.e. a full cylinder, and $R_{i}=1.9 \lambdabar_{a}$, i.e a cylindrical shell with thickness $d=0.1 \lambdabar_{a}$. 
For comparison the Casimir-Polder energy for planar structures, a halfspace and slab of thickness $d$, all comprised of the same material, are also represented. The order of the Gauss-Laguerre and Gauss-Jacobi quadrature is fixed at $N_{\eta} = 10$ and $N_{\zeta} = 30$. For the direct summation, the number of terms is fixed at $N=100$, while the MDL quadrature uses $N=10$ terms. 
}
\label{fig:casimir-polder-hollow-core-cylinder-dielectric-distance-dependence-mdl-comparison}
\end{figure*}
Finally, we need to consider the evaluation of the infinite series over $m$ in \cref{eq:cp-interaction-dimensionless-text}. 
In the thin-wire limit $s\ll 1$, the summands decay rapidly with increasing order such that a direct summation is close to optimal (see also \cref{subsec:thin-wire-T-zero}). 
In the opposite limit of $s\gg 1$, it is useful to analyze the large-order behavior of the summands, which is again dominated by the behavior of the kernel function $\mathcal{K}_{m}(\eta,s)$.  
To this end, we consider the corresponding asymptotic expansions of the modified Bessel functions for fixed argument $\eta$ as $m\to\infty$~\cite{Abramowitz71}, i.e.
\begin{equation}
I_m(x) 
\sim 
\frac{1}{\sqrt{2\pi m}} 
\left(\frac{\ee x}{2 m}\right)^{m}
\;\text{and}\;
K_m(x) 
\sim 
\sqrt{\frac{\pi}{2m}}
\left(\frac{\ee x}{2 m}\right)^{-m}
~.
\end{equation}
As a consequence, for sufficiently large values of $m$ \cref{eq:cp-kernel-function} scales as
\begin{equation}
\begin{aligned}
\mathcal{K}_{m}(\eta, s)
&\sim
\eta^{3}
\frac{1}{2 m} 
\left( 1 + \frac{1}{s} \right)^{-2m}
\\
&=
\eta^{3}
\frac{1}{2 m}
\ee^{-2 m \log{\left(1+\frac{1}{s}\right)}}
~.
\end{aligned}
\label{eq:cp-kernel-function-large-m-asymptotics}
\end{equation}
From this expression, we conclude that for large $m$, the summands decrease exponentially with a decay rate $\mu = 2 \log{(1+1/s)}$. 
Since ${\mu \sim 2 L / R_{0}}$ for ${s \gg 1}$, the decay rate becomes very small for large aspect ratios. 
This results in a slowly convergent sum, confirming our earlier argument based on the nonsuitability of a cylindrical basis expansion in the planar limit. 
A central observation in our reasoning is that -- in the limit of \cref{eq:cp-kernel-function-large-m-asymptotics} -- \cref{eq:cp-interaction-dimensionless-text} has the form of an exponentially decaying bosonic Matsubara sum~\cite{Lifshitz1980}. 
This is functionally the same behavior as observed in the treatment of finite-temperature Casimir forces at imaginary frequencies~\cite{Brevik2005,Rodriguez2011a,Xu2021}. 
More precisely, within this analogy, the radius $R_{0}$ effectively plays the role of an inverse temperature. 
An infinite series of this type can be efficiently evaluated numerically employing a recently developed Gaussian summation technique based on modified discrete Laguerre (MDL) polynomials~\cite{Xu2021}. 
Taken together, the strategy outlined above provides a multivariate Gaussian quadrature scheme for the evaluation of the Casimir-Polder interaction in cylindrical geometries, 
which utilizes a carefully crafted $N_{i}$-point fixed quadrature rule in each dimension ($i \in \{\eta,\zeta,m\}$).

To demonstrate the validity of this approach, we depict in \cref{fig:casimir-polder-hollow-core-cylinder-dielectric-distance-dependence-mdl-comparison} the zero temperature Casimir-Polder potential for a Rubidium atom interacting with a hollow-core silica fiber as a function of the ratio $L/R_{0}$ (similar results are achieved for conductors). We compare the MDL approach with the direct term-wise summation in \cref{eq:cp-interaction-dimensionless-text} for a fixed quadrature order of the $\eta$ and $\zeta$ integrals, using $N_{\eta}=10$ and $N_{\zeta}=30$, respectively.  
For $s\gg 1$, where we expect to recover the result for a planar surface, only a few terms of the MDL summation are required for achieving convergence.
In the examples shown in \cref{fig:casimir-polder-hollow-core-cylinder-dielectric-distance-dependence-mdl-comparison}, we can see that the direct sum approach even with $N_{m}=100$ terms does not yield satisfactory results for large $s$, while $N=10$ MDL terms are sufficient to recover the expected planar limit. 
%. 

\section{Zero temperature interaction}
\label{sec:zero-temperature-interaction}

Next, we proceed to validate our numerical approach by a detailed investigation based on an asymptotic analysis of  \cref{eq:cp-interaction-dimensionless-text}. 
In particular, we present a structured approach, which not only recovers results reported in the literature but also complements and generalizes these results, thereby emphasizing special features of the interaction induced by the interplay between material properties and geometry. 
In the following, we analyze the behavior of the Casimir-Polder energy as function of the distance $L$. 
Depending on the material of the structure, the behavior can be quite different (see \cref{fig:casimir-polder-hollow-core-cylinder-dielectric-distance-dependence,fig:casimir-polder-hollow-cylinder-conductors-distance-dependence}). 

Two limiting configurations can be identified: The slab limit, when $L\ll R_{0}$ ($s\gg 1$), and the thin-wire limit when $L\gg R_{0}$ ($s\ll 1$). 
In light of the numerical analyses presented below, for the slab limit here we simply state the asymptotic expression for the potential within the near field limit ($L\ll \lambdabar_{a}$) and when $ L\ll d$. 
In this setting, the slab appears as a semi-infinite bulk and the Casimir-Polder potential behaves as
~\cite{Bostrom2000,Contreras-Reyes09,Dufour13,Buhmann2012,Buhmann2012a}
\begin{equation}
\label{eq:thickslab-nonretarded-limit}
\mathcal{F}
\sim  - \frac{\hbar}{16 \pi^{2} L^{3}}
 \int\limits_{0}^{\infty} \dd{\xi}\,\frac{\frac{\alpha(\ii \xi)}{\epsilon_{0}}}{1+2\Delta^{2}}
~.
\end{equation}
Further, the perfect conductor limit  in \cref{eq:CasimirPolder48} is recovered for $\Delta=0$.
For larger distances exceeding the shell thickness, i.e.\ when $L \gtrsim d$, the slab thickness becomes important even in the nonretarded limit: the interaction decays with a progressively steeper power law in the atom-surface separation $L$. In dielectrics, this effect is particularly pronounced: in the limit $d \ll L \ll \lambdabar_{a}$, the Casimir-Polder potential is reduced by a factor $\sim d/L$ compared with \cref{eq:thickslab-nonretarded-limit}, yielding a scaling $\propto d/L^{4}$~\cite{Contreras-Reyes09,Buhmann2012,Buhmann2012a}.

In the remainder of the manuscript, we concentrate our analytic investigations on the more geometry-related thin-wire limit.
In order to avoid a cluttered presentation, the next section contains the most relevant information of our analysis. Further details are collected in Appendix~\ref{app:asymptotics}.

\subsection{The thin-wire limit}
\label{subsec:thin-wire-T-zero}

In the thin-wire limit ($s \ll 1$), the interaction effectively behaves as that between an atom and a one-dimensional infinitely extended wire. 
In this limit, noting that $1 + s \sim 1$, the integrand in the Casimir-Polder free energy as presented in \cref{eq:cp-interaction-dimensionless-text} deviates significantly from zero only for $\eta \approx 1$. 
This is primarily due to the exponentially decaying nature of the modified Bessel functions incorporated into $\mathcal{K}_{m}(\eta,s)$ (see  \cref{eq:cp-kernel-function}), which effectively eliminates the contributions arising from large $\eta$. 
Consequently, we can write (see \cref{app:green-tensor})
\begin{equation}
\label{eq:kernel-thin-wire}
  \mathcal{K}_{m}(\eta,s)\sim
  \begin{dcases}
    \frac{\eta^{3}}{-\ln(\eta s \tilde{\gamma}_{\mathrm{E}} )} K_0^2(\eta )& m=0
\\
   m\eta^{3}\left(\frac{\eta s}{2}\right)^{2m} \left[\frac{K_m(\eta )}{m!}\right]^{2} & m\neq 0
  \end{dcases}
\end{equation}
where we have defined $\tilde{\gamma}_{\mathrm{E}}\equiv e^{\gamma_{\mathrm{E}}}/2$, and $\gamma_{\mathrm{E}}$ is the Euler–Mascheroni constant. 
We remark that for $m=0$, the previous approximation introduces an artificial divergence for $\eta \sim (s \tilde{\gamma}_{\mathrm{E}})^{-1}\gg 1$ which does not exist in the original definition of $\mathcal{K}_{m}(\eta,s)$. 
Nevertheless, this issue is inconsequential to our final outcomes, as the resulting expressions remain convergent within the validity range of the approximation, where large values of $\eta$ are suppressed (see also \cref{eq:cp-interaction-zero-temperature-simplified}).

Additionally, we have that for $\eta \approx 1$ and $s\ll 1$ the scattering coefficients in \cref{eq:scattering-coefficients-hollow-cylinder} can be further approximated as follows. 
For the case $m=0$, we can write
\begin{align}
\label{eq:r0NN}
r_{0}^{\mathrm{NN}}
&
\sim 
\frac{(\eta s)^{2}\ln(\eta s \tilde{\gamma}_{\mathrm{E}})
}
{(\eta s)^{2}\ln(\eta s \tilde{\gamma}_{\mathrm{E}})- 2\frac{\Delta^{2}}{1-\chi^{2}}
} ~,
\end{align}
while $r_{0}^{\mathrm{MM}}\sim 0$ and $r_{0}^{\mathrm{MN}}\sim 0$. 
Here, we have defined $\chi= R_{i} / R_{0} = 1 - d / R_{0} < 1$.
The function $r_{0}^{\mathrm{NN}}$ has no explicit dependence on the variable $\zeta$, which only implicitly appears in the argument of $\Delta$. 
We notice that, compared to the full cylinder limit ($\chi=0$), the thickness of the cylindrical shell effectively leads to an increase of the relevance of $\Delta$. 
In terms of the characteristic quantities defined in \cref{eq:relative-penetration-depth-materials-zero-frequency} we basically obtain a set of renormalized parameters according to
\begin{equation}
\label{eq:increased-characteristic-lengths}
\Delta_{0}\to\frac{\Delta_{0}}{\sqrt{1-\chi^{2}}},\quad
 \lambdabar_{p}\to\frac{\lambdabar_{p}}{\sqrt{1-\chi^{2}}}, \quad
 \lambdabar_{D}\to\frac{\lambdabar_{D}}{1-\chi^{2}}~,
\end{equation}
pointing to larger values of these quantities for smaller $d$ and indicating a larger impact of the shell thickness on ohmic conductors than on non-dissipative conductors (see below).
Since the behavior of the denominator of \cref{eq:r0NN} is going to play a central role in the subsequent analysis, we find it convenient to introduce the function 
\begin{equation}
\label{eq:functionW}
\mathcal{W}\left[x\right]\equiv \frac{-x}{W_{-1}(-x)}~,
\end{equation}
where $W_{-1}(x)$ is the second branch of the Lambert-W function.
The expression in \cref{eq:functionW} corresponds to  the smallest solution of the equation $y \ln(y)=-x$ with $x>0$ (if it exists). 
The function $\mathcal{W}\left[x\right]$ monotonically increases with $x$ and is defined up to $\ln(x)=-1$, where we have that $\ln(\mathcal{W})=-1$. 
For $\ln(x)>-1$ we always have that $x> - y \ln(y)$ and no solution exists.

For $m\neq 0$, in the thin-wire limit the scattering coefficients have a very different form. 
They are given by
\begin{subequations}
\label{eq:scattering-coeff-thin-wire-mnot0}
\begin{align}
r_{m}^{\mathrm{NN}}&\sim \frac{1-\chi^{2 m}}{1-\frac{\chi^{2 m}}{\left(1+2 \Delta ^2\right)^2}}
\frac{1-\zeta ^2}{1+2 \Delta ^2}~,
\end{align}

\begin{align}
r_{m}^{\mathrm{MM}}
\sim - \frac{1-\chi^{2 m}}{1-\frac{\chi^{2 m}}{\left(1+2 \Delta ^2\right)^2}}
\frac{\zeta ^2}{1+2 \Delta ^2}~,
\end{align}

\begin{align}
r_{m}^{\mathrm{MN}}
&\sim - \frac{1-\chi^{2 m}}{1-\frac{\chi^{2 m}}{\left(1+2 \Delta ^2\right)^2}}
\frac{\zeta \sqrt{1-\zeta ^2}}{1+2 \Delta ^2}~.
\end{align}
\end{subequations}
Notably, they loose all explicit dependence on the variable $\eta$, which only implicitly appears in the argument of the relative penetration depth $\Delta$.
With respect to the full cylinder case ($\chi=0$), the scattering coefficients of the hollow-core cylinder acquire the same nontrivial prefactor which depends on both the shell thickness and the value of $\Delta$. 
Contrary to $m=0$, the expressions in \cref{eq:scattering-coeff-thin-wire-mnot0} have a structure similar to that of the reflection coefficients of a slab~\cite{Sipe81,Oelschlager18}: The term $\chi^{m}$ takes the role of the propagation factor for the multiple reflections within the layer, highlighting how, even in the thin-wire limit, the shell thickness impacts the electromagnetic scattering for $m>0$. 

The formulas in \cref{eq:kernel-thin-wire,eq:r0NN,eq:scattering-coeff-thin-wire-mnot0} are independent of the nature of the material. 
When inserted in the expression for the Casimir-Polder free energy in \cref{eq:cp-interaction-dimensionless-text}, each of the terms in the sum over $m$ in the resulting integrand is a term proportional to a power of $s$ (see in particular \cref{eq:kernel-thin-wire}). 
For $s\ll 1$ the dominant contribution arises from terms related to the orders $m=0$ and $m=1$ (see \cref{eq:cp-interaction-zero-temperature-simplified}).
To determine the asymptotic behavior of the interaction, we need to identify the regions providing the significant contributions to the integrals over $\eta$ and $\zeta$. While the $\eta$-integration is mainly affected by the kernel function  $\mathcal{K}_{m}(\eta,s)$, the polarizability mostly influences the $\zeta$-integration, distinguishing between the nonretarded ($L\ll \lambdabar_{a}$) and the retarded limit ($L\gg \lambdabar_{a}$). 
We note that within the thin-wire limit, the nonretarded region can only be realized if $R_{0}/\lambdabar_{a} \ll 1$. 
This sets an upper bound to the possible values of the radius for which the following considerations are valid. 
Whenever $R_{0}/\lambdabar_{a} \gtrsim 1$, then in the nonretarded region the slab limit is systematically recovered (see \cref{eq:thickslab-nonretarded-limit} and the accompanying discussion above).
Within this constraint, an inspection of \cref{eq:cp-interaction-dimensionless-text} reveals that, given that the polarizability is significantly different from zero for $0\le \xi\lesssim \omega_{a}$ (as per \cref{eq:polarizability-model}),
the integrations are effectively restricted to the regions where $0\le \zeta \eta \lesssim L/\lambdabar_{a}$.
Given that $\mathcal{K}_{m}(\eta,s)$ additionally limits the pertinent values to $\eta \approx 1$, in the nonretarded limit we have that $\zeta \ll 1$. We can then further simplify \cref{eq:cp-interaction-dimensionless-text} by neglecting all explicit dependencies on the variable $\zeta$ with respect to the other terms, both in the integrand and in the scattering coefficients, without altering the implicit dependence via $\alpha$ and $\Delta$. In addition, we can, to a good approximation
extend the upper limit of the $\zeta$-integration to infinity.
Conversely, when retardation effects become relevant, i.e. for $L/\lambdabar_a \gg 1$, since we have in general that $0<\zeta\le 1$, the arguments of $\alpha $ in and $\Delta$ \cref{eq:cp-interaction-dimensionless-text} become small. Specifically, the polarizability can be approximated by its static value and, if $L\gg c/\xi_{0}$, $\Delta$ can be approximated by its low-frequency behavior. 
According to \cref{eq:relative-penetration-depth-materials-zero-frequency}, within the retarded region, we then expect quite different results for the Casimir-Polder interaction, differing not only between conductors and dielectrics but also between ohmic and lossless conductors. 

To see this in detail, in the following we discuss the results of the above approximations for dielectrics separately from those for conductors, in both cases analyzing the nonretarded and retarded limits within the thin-wire approximation, as well as comparing the results to a full numerical evaluation. 

\subsubsection{Dielectrics}
\label{sec:dielectric-hollow-core-cylinder}

One of the most relevant characteristics of dielectric materials, clearly distinguishing them from conductors, is their behavior at low frequencies as highlighted by \cref{eq:relative-penetration-depth-materials-zero-frequency}. 
In particular, at low frequencies we have that $\Delta \to \Delta_{0}>0$, i.e. the relative penetration depth approaches a positive constant that is connected to the dielectric's insulating properties. 
This has a profound effect on the behavior of $r^{\mathrm{NN}}_{0}$ in \cref{eq:r0NN}: if $s$ is sufficiently small (see below) and given that $\eta \approx 1$, we can neglect the logarithmic term in the denominator.
Using the function $\mathcal{W}[x]$ defined in \cref{eq:functionW} we can formulate this condition more precisely as 
\begin{subequations} 
\label{eq:condition-dielectrics}
\begin{equation}
 s \ll  s_{0}\equiv\frac{1}{\tilde{\gamma}_{\mathrm{E}}}\sqrt{\mathcal{W}\left[\left(2\frac{\Delta_{0}}{\sqrt{1-\chi^{2}}}\tilde{\gamma}_{\mathrm{E}}\right)^{2}\right]}
~,
\end{equation}
which sets a lower bound to the distances this approximation is valid for. 
Due to the properties of $\mathcal{W}\left[x\right]$, this condition is fulfilled by all distances compatible with the thin-wire limit if 
\begin{equation}
\ln\left(2\frac{\Delta_{0}}{\sqrt{1-\chi^{2}}}\tilde{\gamma}_{\mathrm{E}}\right)^{2}\ge -1~,
\end{equation}
\end{subequations}
distinctly illustrating how the value of the shell thickness modifies the validity of the asymptotic descriptions we shall now derive. 

Within the above approximation, an examination of \cref{eq:cp-interaction-dimensionless-text} indicates that the terms corresponding to $m=0$ and $m=1$ both scale as $s^{2}$, while higher values of $m$ yield terms proportional to $s^{2m}$ and can therefore be disregarded.
If $R_{0}\ll L\ll \lambdabar_{a}$, the nonretarded region overlaps with the thin-wire limit and as discussed above we can neglect the explicit dependencies on the variable $\zeta$, both in the integrand and in the scattering coefficients.
With this simplification $r^{\mathrm{MM}}_{m}$ and $r^{\mathrm{MN}}_{m}$ do not contribute to the interaction and we are left only with $r^{\mathrm{NN}}_{m}$ for $m=0,1$ (see \cref{eq:cp-interaction-zero-temperature-simplified}). 
Reintroducing the integration variable $\xi=\zeta \eta s c/R_{0}$, the integration over $\eta$ of the resulting expression can be performed analytically leading to
\begin{equation}
\label{eq:cp-hollow-core-nonretarded}
\mathcal{F}
\sim
- 
\frac{9\hbar}{512 \pi } \frac{R^{2}_{0}}{L^5}
\int\limits_{0}^{\infty} \dd{\xi} \, \frac{\alpha(\ii \xi)}{\epsilon_{0}}
\left[
\frac{1-\chi^{2}}
{\Delta(\ii \xi)^{2}} 
  +3
\frac{ \frac{1-\chi^{2}}{1-\frac{\chi^{2}}{\left[1+2 \Delta(\ii \xi)^2\right]^2}}}{1+2 \Delta(\ii \xi)^2}
\right]~.
\end{equation}
In this expression, the effects of the surface curvature are evident.
Within the nonretarded limit, if $R_{0}\ll L$ the curvature leads to a Casimir-Polder free energy that roughly scales $\propto R^{2}_{0}(1-\chi^{2})/L^{5}$ which is faster than $\propto d/L^{4}$ observed for a thin dielectric slab with the same thickness of the cylindrical shell. 
Note that for dielectrics, the shell thickness alters the strength of the interaction without altering the scaling. 
The expression in \cref{eq:cp-hollow-core-nonretarded} vanishes for $d \to 0$ and/or $R_{0}\to 0$, while the maximum value is attained for $\chi=0$. In other words, the full cylinder~\cite{Nabutovskii1979, Barash1989} is recovered as a particular case of our result. 
The two terms in the square brackets describe the different contributions of the two aforementioned values for the order $m$. Quantitatively, their contribution is similar for $\Delta\gg 1$ but they behave differently in the opposite limit.
The interaction is stronger for small values of $\Delta$, corresponding to a lower penetration depth. In this limit, the largest contribution  arises from the term associated with the $m=0$ term.

We further note that the appearance of the factor $1-\chi^{2}$ in \cref{eq:cp-hollow-core-nonretarded} can be justified from dimensional analysis. 
It amounts to considering the cross-sectional area $\propto R^{2}_{0} - R^{2}_{i}$ of the hollow cylinder as can be seen by writing
\begin{equation}
\label{eq:dimensional-analysis}
\frac{R^{2}_{0}}{L^5}(1-\chi^{2})=\frac{R_0^2-R_i^2}{L^5}=\frac{1}{L^{3}}2\frac{d}{L}\frac{R_{\mathrm{av}}}{L},
\end{equation}
where we have used again $d=R_{0}-R_{i}$ and we have defined the average radius $R_{\mathrm{avg}}=(R_{0}+R_{i})/2$. The first term on the r.h.s. of the above equation, i.e. $L^{-3}$, is the typical nonretarded behavior obtained for planar semi-infinite structures (see \cref{eq:CasimirPolder48,eq:thickslab-nonretarded-limit}~\cite{Intravaia2011,Buhmann2012,Buhmann2012a}). 
In addition, one has to take into account that for dielectrics this trend is corrected by the shell thickness through the dimensionless factor $d/L$, i.e. the second term on the r.h.s. of \cref{eq:dimensional-analysis} (see also \cref{eq:thickslab-nonretarded-limit} and the related discussion). 
Finally, the curvature of the surface further reduces the geometrical extent of the structure introducing another factor $R_{\mathrm{avg}}/L$, which gives rise to the last term on the r.h.s. of \cref{eq:dimensional-analysis}. 

Now, moving on to the retarded regime, we can replace $\alpha\to\alpha_{0}$ in \cref{eq:cp-interaction-dimensionless-text}
and, considering that $\zeta\le 1$ and that the relevant contributions arise from the region $\eta\approx 1$ for $L\gg c/\xi_{0}$, the function $\Delta$ can be approximated by its low frequency behavior according to \cref{eq:relative-penetration-depth-materials-zero-frequency}. 
Within this approximation, the resulting integrals (see \cref{eq:cp-interaction-zero-temperature-simplified}) can be analytically evaluated leading to
\begin{equation}
\mathcal{F}\sim
- 
\frac{\hbar c}{60 \pi^2 } \frac{\alpha_{0}}{\epsilon_{0}}\frac{R_{0}^{2}}{L^{6}}\left[7\frac{1-\chi^{2}}{\Delta_{0}^{2}}+32\frac{\frac{1-\chi^{2}}{1-\frac{\chi^{2}}{\left[1+2 \Delta_{0}^2\right]^2}}}{1+2 \Delta_{0} ^2}\right] 
~,
\label{eq:cp-hollow-core-retarded-dielectric}
\end{equation}
which again generalizes the result for the full cylinder~\cite{Barash1989}.
Similar to the nonretarded case, the first term in the square brackets above originates from $r^{\mathrm{NN}}_{0}$ in \cref{eq:r0NN}, while the second term results from the collective contribution of all three scattering coefficients $r^{\mathrm{ij}}_{1}$ in Eqs.~\eqref{eq:scattering-coeff-thin-wire-mnot0}.
The power-law behavior $\propto L^{-6}$ can be understood in terms of dimensional analysis as in \cref{eq:dimensional-analysis}, with the sole difference that in the retarded region for planar structures the interaction scales as $\propto L^{-4}$ (see  \cref{eq:CasimirPolder48})~\cite{Casimir1948,Intravaia2011,Buhmann2012,Buhmann2012a}.
%%%%%%%%%%%%%%%%%%%%%%%%%%%%%%%%%%%%%%%%%%%%%%%%%%%%%%%%%%
\begin{figure}
  \centering
  \includegraphics[width=\linewidth]{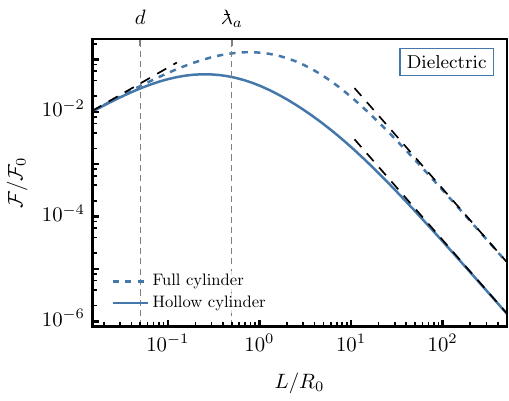}
  \caption{
 Zero-temperature Casimir-Polder potential for a Rubidium atom interacting with a hollow-core silica fiber of radius $R_{0}=2\lambdabar_{a}$ (see \cref{fig:casimir-polder-hollow-core-cylinder-dielectric-distance-dependence-mdl-comparison} for the corresponding parameters). The potential is represented as a function of the atom-structure separation for $\chi=0$ (full cylinder) and $\chi=0.95$ ($d=0.1\lambdabar_{a}$). All expressions are normalized by the retarded interaction between an atom and a perfectly reflecting surface, $\mathcal{F}_{0}$ in \cref{eq:CasimirPolder48}. 
The different curves highlight how the free energy is affected by the material properties, the curvature and the thickness of the shell. At short atom-cylinder separations, all expressions reproduce the case of an atom interacting with a dielectric half space, \cref{eq:thickslab-nonretarded-limit}. As the distance increases, the full cylinder continues to follow the behavior of the halfspace $\propto L^{-3}$ (dotted line), while the dielectric cylindrical shell has the same trend as a dielectric slab of the same thickness with a power law between $L^{-3}$ and $L^{-4}$ (see the comment below \cref{eq:thickslab-nonretarded-limit}). For large distances, according to \cref{eq:cp-hollow-core-retarded-dielectric}, both the full cylinder and the cylindrical shell scale as $\propto L^{-6}$ (black dashed lines), featuring a decay which is faster than the interaction with a halfspace ($\propto L^{-4}$) and with a slab ($\propto L^{-5}$~\cite{Buhmann2012,Buhmann2012a}).
The curves illustrate the modification in the strength of the interaction introduced by the shell's thickness. In agreement with \cref{eq:cp-hollow-core-retarded-dielectric}, the maximum strength is reached for $\chi=0$ (full cylinder).} 
\label{fig:casimir-polder-hollow-core-cylinder-dielectric-distance-dependence}
\end{figure}
%
%%%%%%%%%%%%%%%%%%%%%%%%%%%%%%%%%%%%%%%%%%%%%%%%%%%%%%%%%%
Employing the full numerical evaluation of the Casimir-Polder interaction (\cref{sec:numerical-evaluation}), we depict in 
\cref{fig:casimir-polder-hollow-core-cylinder-dielectric-distance-dependence} 
the results for a hollow-core fiber ($\chi = 0.95$) over a wide range of the atom-surface separation.
To highlight the impact of the shell thickness on the behavior of $\mathcal{F}$, we additionally report the interaction for a full cylinder of the same radius ($\chi=0$). 
The dielectric fiber is described by the Drude-Lorentz model presented in \cref{eq:permittivity-models}, with parameters corresponding to silica (see \cref{fig:casimir-polder-hollow-core-cylinder-dielectric-distance-dependence-mdl-comparison}). 
These parameters yield a $\Delta_{0}$ that fulfills the condition in \cref{eq:condition-dielectrics} for all distances compatible with the thin-wire limit.
The numerical analysis confirms that at very short distances the interaction tends to the nonretarded Casimir-Polder potential of an atom in front of a dielectric halfspace, scaling as $L^{-3}$. 
For the parameters considered in \cref{fig:casimir-polder-hollow-core-cylinder-dielectric-distance-dependence}, as the distance increases and becomes larger than the shell thickness $d$, the full cylinder continues to follow the halfspace behavior \cite{Intravaia2011,Buhmann2012,Buhmann2012a}.
Conversely, given the relatively small value of $d$ (see \cref{fig:casimir-polder-hollow-core-cylinder-dielectric-distance-dependence}), the dielectric hollow-core fiber has the same trend as a dielectric slab with the same thickness and the corresponding power law in between $L^{-3}$, typical for the halfspace, and $L^{-4}$ (see the discussion below \cref{eq:thickslab-nonretarded-limit}). 
Given that $R_{0}/\lambdabar_{a}>1$, the asymptotic expression presented in \cref{eq:cp-hollow-core-nonretarded} is not applicable in this scenario, but we have verified its consistency with a numerical evaluation using other parameters (not shown). 
In \cref{fig:casimir-polder-hollow-core-cylinder-dielectric-distance-dependence}, we see that for $L/R_{0}\gg 1$, the numerical evaluation recovers the asymptotic expressions given in \cref{eq:cp-hollow-core-retarded-dielectric}. 
In comparison to its hollow-core counterpart, the full cylinder shows a stronger Casimir-Polder interaction, in agreement with \cref{eq:cp-hollow-core-retarded-dielectric}. 

\subsubsection{Conductors}
\label{subsubsec:thin-wire-T-zero-conductors}

An important distinguishing feature between dielectrics and conductors is their behavior at low frequencies. 
In particular, the asymptotic expressions in \cref{eq:cp-hollow-core-nonretarded,eq:cp-hollow-core-retarded-dielectric} are derived by assuming that at low frequencies,  
$\Delta$ does not vanish.
They manifestly do not work for conductors, where conduction electrons can lead to $\Delta\to 0$: In this case, the first term in the square brackets of these equations diverges~\cite{Barash1989}.
Reanalyzing the contribution of the scattering coefficients in the limit $s\ll 1$, we notice that for $m > 0$, all the scattering coefficients are well-behaved for $\Delta\to 0$, enabling considerations which are similar to those previously formulated for dielectrics.
Conversely, the most immediate consequence of $\Delta\to 0$ is that the logarithmic term in the denominator of $r^{\mathrm{NN}}_{0}$ in \cref{eq:r0NN} can no longer be neglected, and can in some cases even constitute the dominant behavior. 
This implies that, differently from the dielectric case, for conductors the leading order in the limit $s\ll 1$ is provided by only the $m=0$ term in \cref{eq:cp-interaction-dimensionless-text}, solely stemming from the $\mathrm{NN}$-polarization. 

We start by considering the nonretarded regime  $R_{0}\ll L\ll \lambdabar_{a}$. As mentioned above, the range of validity of the thin-wire limit is in this case restricted by a lower bound on $R_{0}/\lambdabar_{a}\ll s \ll 1 $.
As in the previous sections, we can remove all explicit dependencies on the variable $\zeta$ in the corresponding integration and extend the upper bound of the integral to infinity. 
Since the only relevant contribution comes from $r^{\mathrm{NN}}_{0}$, the resulting integrand has the same denominator as \cref{eq:r0NN}.  
With respect to $\eta^{4} [1 + \mathbb{k}^{2}_0(\eta)]K_0^2(\eta)>0$ appearing in the numerator of the integrand (see \cref{eq:conductor-nonretarded}), the remaining part of the expression varies slowly as a function of $\eta$.
In general, an expression of the form $\eta^{j}K_0^2(\eta)$ peaks at $\eta=\eta_{[j]}\approx (j-1)/2$, which provides the dominant contribution to the $\eta$-integration. 
Using a saddle point approximation (see \cref{app:asymptotics}), we can perform the corresponding integration obtaining
\begin{equation}
\label{eq:conductor-nonretarded-start}
\mathcal{F}\sim
- 
\frac{9\hbar}{512 \pi } \frac{R_{0}^{2}}{L^5}
\int\limits_{0}^{\infty} \dd{\xi}\; \frac{ \frac{\alpha\left(\ii \xi\right)}{\epsilon_{0}} }
{\frac{\Delta^{2}\left(\ii \xi \right)}{1-\chi^{2}}-\frac{(\eta_{[4]} s)^{2} }{2}\ln(\eta_{[4]} s \tilde{\gamma}_{\mathrm{E}})}
~.
\end{equation}
We further notice that the polarizability truncates the relevant contribution to the integral at $\xi\sim \omega_{a}$, and, given that $\Delta$ is a monotonically growing function, $\Delta(\ii\omega_{a})$ is effectively its largest relevant value. 
Using the definition in \cref{eq:functionW}, if the condition
\begin{equation}
\label{eq:conductor-condition-nonretarded}
 \frac{1}{\eta_{[4]}\tilde{\gamma}_{\mathrm{E}}} \sqrt{\mathcal{W}\left[ \left(2 \frac{\Delta\left(\ii \omega_{a} \right)}{\sqrt{1-\chi^{2}}}\tilde{\gamma}_{\mathrm{E}}\right)^{2}\right]}  \ll \frac{R_{0}}{\lambdabar_{a}} \ll  1
\end{equation}
is fulfilled, we can neglect the first term in the denominator of \cref{eq:conductor-nonretarded-start} with respect to the term containing the logarithm and write
\begin{align}
\label{eq:superconductor-nonretarded-1}
\mathcal{F}
&\sim
- 
\frac{9\hbar}{256 \pi } \frac{\frac{1}{\eta_{[4]}^{2} }\int\limits_{0}^{\infty} \dd{\xi}\; \frac{\alpha\left(\ii \xi\right)}{\epsilon_{0}}}
{L^3\ln\left( \frac{\frac{L}{R_{0}}}{\eta_{[4]}  \tilde{\gamma}_{\mathrm{E}}}\right)}~.
\end{align}
We see that, compared to dielectrics, the conductors' Casimir-Polder potential exhibits a different behavior as a function of the distance $L$. 
Nonetheless, here the change is not a modification of the exponent of the power law: it is instead marked by the appearance of a logarithm. 
Notice that the previous expression does not contain any material parameters: They only implicitly appear in the condition in \cref{eq:conductor-condition-nonretarded}. 
In fact, the expression in \cref{eq:superconductor-nonretarded-1} is equivalent to considering $r_{0}^{\mathrm{NN}}=1$, corresponding to the result that would have been obtained for a perfectly conducting cylinder: 
Neglecting $\Delta$ is indeed equivalent to assuming it is vanishing.

Material parameters are more important when, over the interval $0<\xi\lesssim \omega_{a}$, the term containing $\Delta$ in the denominator of \cref{eq:conductor-nonretarded-start} becomes larger in absolute value than the logarithmic term. 
If we denote as $\xi_{L}$ the frequency for which this happens, then $\xi_{L}$ is determined by solving
\begin{equation}
\label{eq:conditionxiL}
\frac{\Delta^{2}\left(\ii \xi \right)}{1-\chi^{2}}=-\frac{(\eta_{[4]} s)^{2} }{2}\ln(\eta_{[4]} s \tilde{\gamma}_{\mathrm{E}})~.
\end{equation}
The value of $\xi_{L}$ depends on the system's geometry and, for $s\ll 1$, it decreases with $s$.
If $\xi_{L}<\omega_{a}$, the dominant contribution to the frequency integral is not determined by the polarizability but rather by the denominator of the integrand in \cref{eq:conductor-nonretarded-start}, thereby modifying the Casimir-Polder potential in terms of the system's geometry and material composition.
More specifically, if we consider the case $\xi_{L}\ll\xi_{0}$, the behavior of $\Delta$ can be described by \cref{eq:relative-penetration-depth-materials-zero-frequency}. 
For lossless conductors we can then approximate the polarizability with its static value, while the remaining integral can be evaluated analytically to give
\begin{align}
\label{eq:superconductor-nonretarded-2}
\mathcal{F}
&\sim
- 
\frac{9\hbar c}{2^{10}} \frac{R_{0}}{L^4}
\frac{\frac{ \alpha_{0}}{\epsilon_{0}\eta_{[4]}}\frac{\sqrt{1-\chi^{2}}}{\lambdabar_{p}}}{\sqrt{2\ln\left( \frac{\frac{L}{R_{0}}}{\eta_{[4]}\tilde{\gamma}_{\mathrm{E}}} \right)}}~.
\end{align}
When contrasted with \cref{eq:superconductor-nonretarded-1}, we can clearly see the impact of the material properties on the free energy's scaling with respect to $L$, where the exponent of the power law has now decreased by one. 
In the case of the ohmic conductor, the polarizability function has to be considered in its full form as in \cref{eq:conductor-nonretarded-start}. 
Nevertheless, within the conditions set by the thin-wire limit, we have 
\begin{align}
\label{eq:conductor-nonretarded}
\mathcal{F}
&\sim
- 
\frac{9\hbar c}{512 \pi }
\frac{\alpha_{0}}{\epsilon_{0}}\frac{1-\chi^{2}}{\lambdabar_{D}} \frac{R_{0}^{2}}{L^5}\ln\left(\frac{2\frac{\frac{\lambdabar_{D}}{1-\chi^{2}}}{\lambdabar_{a}}\left( \frac{\frac{L}{R_{0}}}{\eta_{[4]}}\right)^{2} }{\ln\left(\frac{\frac{L}{R_{0}}} {\eta_{[4]}\tilde{\gamma}_{\mathrm{E}}}\right)}\right)~.
\end{align}
We notice the yet different behavior of $\mathcal{F}$ with respect to \cref{eq:superconductor-nonretarded-1,eq:superconductor-nonretarded-2}, highlighting how the interplay between geometry and material properties can affect the behavior of the Casimir-Polder free energy in multiple ways.

A few additional comments about $\xi_{L}$ are in order. 
Since $\xi_{L}\ll \xi_{0}$, using \cref{eq:conditionxiL}, we need to have
\begin{align} 
\label{eq:constraint-s}
\frac{R_{0}}{\lambdabar_{a}}\ll
s\ll   \frac{1}{\eta_{[4]}\tilde{\gamma}_{\mathrm{E}}} \sqrt{\mathcal{W}\left[ \left(2 \frac{\Delta\left(\ii \xi_{0} \right)}{\sqrt{1-\chi^{2}}}\tilde{\gamma}_{\mathrm{E}}\right)^{2}\right]} ~,
\end{align}
which sets an upper bound to the value of $s$ for which the approximated expressions are viable. 
As for dielectrics, the upper bound of this condition is automatically fulfilled when the logarithm of the argument of $\mathcal{W}$ is equal to minus one.
In particular, the specific relative permittivity models given in \cref{eq:permittivity-models} require $\xi_{L} \ll \min \{\omega_{a},\omega_{p}/\sqrt{\epsilon_{\infty}-1}\}$ for a superconducting material and $\xi_{L} \ll \min\{ \gamma , \omega_{a}\}$ for an ohmic conductor. 
For the latter, the constraints imposed by Eqs.~\eqref{eq:constraint-s} might be difficult to fulfill in relation to the typical values of its parameters. The required values of $\chi$ are in this case very close to one, corresponding to very thin shells and/or small values of $s$ which are not compatible with the nonretarded region's lower bound $s\gg R_{0}/\lambdabar_{a}$.

Let us now consider the interaction for large atom-surface separations, i.e. in the retarded regime, where $L\gg \lambdabar_{a}$.
Again, in the thin-wire limit, the dominant contribution to \cref{eq:cp-interaction-dimensionless-text} is given by the $m=0$ term. For these distances, the polarizability can be approximated by its static value and the behavior of the Casimir-Polder free energy is asymptotically described by
\begin{equation}
\mathcal{F}
\sim
- 
\frac{\hbar c}{8 \pi^3} \frac{\frac{\alpha_{0}}{ \epsilon_{0}}}{L^4}
\int\limits_{0}^{\infty} \dd{\eta}
\int\limits_0^{1} \dd{\zeta} \; 
 \frac{\frac{1 + (1-\zeta^{2}) \mathbb{k}^{2}_0(\eta)}{\sqrt{1-\zeta^2}}\eta^{3}K_0^2(\eta ) }
{\frac{\Delta^{2}\left(\ii\zeta\frac{c}{R_{0}}\eta s\right)}{(\eta s)^{2}(1-\chi^{2})}-\frac{\ln(\eta s \tilde{\gamma}_{\mathrm{E}})}{2} }
~.
\label{eq:conductor-retarded-start}
\end{equation}
We focus again on the frequency region where we can approximate the function $\Delta$ according to \cref{eq:relative-penetration-depth-materials-zero-frequency}. Taking into account that $\eta\approx1$ and $\zeta\le 1$, this implies that $ s\ll R_{0}\xi_{0}/c$. 

For lossless conductors, the first term in the denominator of \cref{eq:conductor-retarded-start} becomes proportional to $\zeta^{2}$ and loses its dependence on $\eta s$, which is only contained in the logarithmic term. 
Employing a similar strategy as in the calculation of the nonretarded limit above, we notice that the numerator of the above integrand is positive and peaks around $\eta \approx \eta_{[3]}$ (the effect of $\zeta$ on the peak of the numerator of \cref{eq:conductor-retarded-start} is negligible). 
We can then again perform the integral over $\eta$ using the saddle-point approximation (see App. \ref{app:thin-wire}).   
As above, the main contribution to the integral in the previous expression arises from the values for which the term containing the function $\Delta$ in the denominator of the integrand is smaller than or of the order of the logarithmic term, leading to constraints similar to those of \cref{eq:constraint-s}. 
The resulting expression contains an integral over $\zeta$ which can be calculated analytically (see \cref{superconductor-retarded-app}), leading to
\begin{align}
\label{eq:superconductor-retarded-text}
\mathcal{F}
&\sim-\frac{\hbar c}{8 \pi^2} \frac{\frac{\alpha_{0}}{\epsilon_{0}}}{L^4\ln\left(\frac{\frac{L}{R_{0}}}{\eta_{[3]} \tilde{\gamma}_{\mathrm{E}}}\right) }
\frac{1-\frac{\frac{2}{3}}{1+\sqrt{1+\frac{\frac{2}{1-\chi^{2}}\left(\frac{\lambdabar_{p}}{R_{0}}\right)^{2}}{\ln\left(\frac{\frac{L}{R_{0}}}{\eta_{[3]} \tilde{\gamma}_{\mathrm{E}}}\right)}}}}{\sqrt{1+\frac{\frac{2}{1-\chi^{2}}\left(\frac{\lambdabar_{p}}{R_{0}}\right)^{2}}{\ln\left(\frac{\frac{L}{R_{0}}}{\eta_{[3]} \tilde{\gamma}_{\mathrm{E}}}\right)}}}
~.
\end{align}
We remark that, under the condition%
\begin{equation}
\label{eq:condition-superconductor-retarded}
-\frac{\frac{2}{1-\chi^{2}}\left(\frac{\lambdabar_{p}}{R_{0}}\right)^{2}}{\ln\left(\eta_{[3]} \tilde{\gamma}_{\mathrm{E}}s\right)}\ll 1 \quad
\Rightarrow
s\ll\; s_{p}\equiv \frac{e^{-\frac{2}{1-\chi^{2}}\left(\frac{\lambdabar_{p}}{R_{0}}\right)^{2}}}{\eta_{[3]}\tilde{\gamma}_{\mathrm{E}}}~,
\end{equation}
the expression in \cref{eq:superconductor-retarded-text} approaches
\begin{equation}
\label{eq:perfect-conductor-retarded}
\mathcal{F}\sim
-\frac{\hbar c}{12 \pi^2 } \frac{\frac{\alpha_{0}}{\epsilon_{0}}}{L^4\ln\left(\frac{\frac{L}{R_{0}}}{\eta_{[3]} \tilde{\gamma}_{\mathrm{E}}}\right) }
~.
\end{equation}
Similar to \cref{eq:superconductor-nonretarded-1}, the expression in \cref{eq:perfect-conductor-retarded} is independent of any material properties and indeed represents the limit for $L\gg R_{0}$ of the Casimir-Polder interaction with a perfectly conducting cylinder \cite{Barash1989}. 
According to the condition in \cref{eq:condition-superconductor-retarded}, the thinner the shell the larger is the distance for which the perfect conductor limit in \cref{eq:perfect-conductor-retarded} is valid, indicating that, within the thin-wire limit, the interaction is mostly described by \cref{eq:superconductor-retarded-text}. Indeed, $s_{p}$, constraining the range of validity of the previous expressions, exponentially decreases with the inverse thickness of the cylindrical shell.
This is consistent with the fact that thinner shells effectively increase the penetration depth (see \cref{eq:increased-characteristic-lengths}), enhancing the differences between perfect and non-dissipative conductor: The term containing $\Delta$ in the denominator of \cref{eq:conductor-retarded-start}, providing information about the material parameters, becomes more relevant.

For regular conducting materials, as soon as their ohmic behavior becomes relevant, we obtain a quite different result. Indeed, using the corresponding low-frequency approximation for $\Delta$ from \cref{eq:relative-penetration-depth-materials-zero-frequency} in the denominator of the integrand of \cref{eq:conductor-retarded-start}, the term containing the function $\Delta$ now scales as $\zeta/(\eta s)$. 
The extra dependence on $\eta s$ slightly modifies the peak of the numerator which is again located around $\eta_{[4]}$. Proceeding as above, both the integral in $\zeta$ and $\eta$ can be performed analytically (see \cref{eq:conductor-retarded-start-app}). 
The resulting expression can be simplified under the condition
\begin{subequations}
\label{condition-conductor-retarded-full}
\begin{equation}
\begin{gathered}
\label{condition-conductor-retarded}
-\frac{(\eta_{[4]} s)\ln\left(\eta_{[4]} s \tilde{\gamma}_{\mathrm{E}}\right)}{ \frac{2}{1-\chi^{2}}\frac{\lambdabar_{D}}{R_{0}}}\ll 1
\\
\Rightarrow s\ll s_{D}\equiv  \frac{1}{\eta_{[4]}\tilde{\gamma}_{\mathrm{E}}} \mathcal{W}\left[\frac{2} {1-\chi^{2}}\frac{\lambdabar_{D}}{R_{0}}\tilde{\gamma}_{\mathrm{E}}\right]~.
\end{gathered}
\end{equation}
As above, due to the properties of $\mathcal{W}[x]$,  this condition is fulfilled for all $s\ll 1$ if 
\begin{equation}
\label{eq:condition-conductor-retarded-max}
\ln\left(\frac{2} {1-\chi^{2}}\frac{\lambdabar_{D}}{R_{0}}\tilde{\gamma}_{\mathrm{E}}\right)\ge -1~,
\end{equation}
\end{subequations}
showing once again how the shell thickness can be used to tailor the distance regime over which the approximation is valid.
If $s\ll s_{D}$ the Casimir-Polder free energy takes the form
\begin{equation}
\mathcal{F}
\sim 
- \frac{9\hbar c}{2^{12} \pi} \frac{\alpha_{0}}{\epsilon_{0}}  \frac{1-\chi^{2}}{\lambdabar_{D}}\frac{R_{0}^{2}}{L^{5}}
\left[
  8\ln\left(
    \frac{ 4\frac{\frac{\lambdabar_{D}}{1 - \chi^2}}{R_{0}}\frac{\frac{L}{R_{0}}}{\eta_{[4]}}
    }
    {\ln\left( \frac{\frac{L}{R_{0}}}{\eta_{[4]}\tilde{\gamma}_{\mathrm{E}}} \right)}
  \right)
  -5
\right]~.
\label{eq:conductor-retarded}
\end{equation}
\Cref{eq:conductor-retarded} differs in many aspects from \cref{eq:superconductor-retarded-text}, even more from the perfect conductor limit in \cref{eq:perfect-conductor-retarded} and has common traits with \cref{eq:conductor-nonretarded}. 
First of all, as in the case of dielectrics, the strength of the Casimir-Polder interaction scales as the cross-sectional area of the hollow-core cylinder (see \cref{eq:dimensional-analysis}). 
The prefactor is large for small values of $\lambdabar_{D}$ and, notably, the interaction falls off faster with $L$ than in \cref{eq:superconductor-retarded-text}. 
Furthermore, as in \cref{eq:conductor-nonretarded}, the first term inside the square brackets features a peculiar double logarithmic behavior, which is not present in the superconducting case and can be modified as a function of the shell thickness.
Consistent with Ref.~\cite{Barash1989}, the double logarithmic behavior persists in the case of the full cylinder $\chi=0$. 
Note that $s_{D}$ monotonically grows as the inverse of the shell thickness, indicating that for thinner shells, the behavior of the Casimir-Polder potential for the ohmic conductor becomes visible at shorter distances.
In other words, within the thin-wire limit, the range of validity of \cref{eq:conductor-retarded} increases for smaller values of $d$. Again, also in relation to \cref{eq:increased-characteristic-lengths}, this can be understood in terms of a larger effective diffusion length which is equivalent to a more pronounced ohmic behavior.
%%%%%%%%%%%%%%%%%%%%%%%%%%%%%%%%%%%%%%%%%%%%%%%%%%%%%%%%%%
%
\begin{figure*}
\subfloat[]{
  \includegraphics[width=0.99\columnwidth]{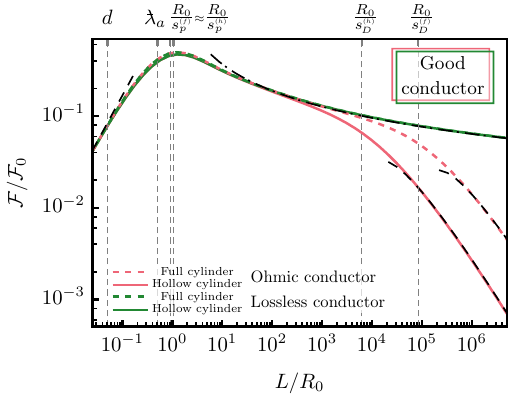}
}
\hspace{0.02\columnwidth}
\subfloat[]{
  \includegraphics[width=0.99\columnwidth]{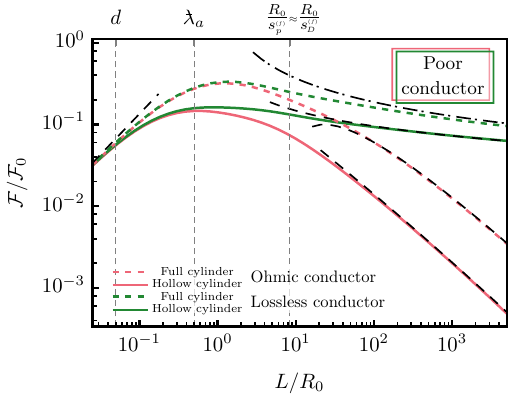}
}
\caption{
The Casimir-Polder potential at $T=0$ for a Rubidium atom interacting with a cylindrical shell comprised of a) a good conductor and b) a poor conductor, respectively. The external radius is chosen as $R_0=2\lambdabar_a$ and the values $\chi=0$ (full cylinder) and $\chi=0.95$ (hollow cylinder) are considered. 
All expressions are normalized with respect to the retarded limit of the free energy for an atom interacting with a perfectly conducting plane (see \cref{eq:CasimirPolder48}). The potential is represented as a function of the atom-cylinder separation.
The different curves highlight how the interaction is affected by the material properties, the curvature and the thickness of the shell. 
\textit{Good conductor}: In the expression for the Drude model given in \cref{eq:permittivity-models}, we use $\epsilon_{\infty}=1$, $\omega_{p}=\SI{9}{\eV}$ and $\gamma=10^{-2}\unit{\eV}$, which correspond to a noble metal like gold or silver~\cite{Behunin14a}. For these parameters, within the nonretarded limit for $L\ll d$ the potential behaves accordingly as $\propto L^{-3}$, in agreement with \cref{eq:thickslab-nonretarded-limit}. At larger distances, the impact of the shell thickness is initially marginal.
We have $s_{p}\approx 1$ for both the full ($s^{(f)}_{p}= 1.105$) and the hollow-core ($s^{(h)}_{p}= 0.954$) cylinder, indicating that, within the thin-wire limit, the interaction with the non-ohmic conductor recovers the perfect conductor limit (dash-dotted line).
Conversely, for an ohmic conductor $s^{(f)}_{D}= 0.12 \times 10^{-4}$ for $\chi=0$ and $s^{(h)}_{D}= 1.4 \times 10^{-4}$ for $\chi=0.95$, highlighting where the interaction starts to feature the behavior described by \cref{eq:conductor-retarded} (black dashed lines). 
\textit{Poor conductor}: The material is described by $\epsilon_{\infty}=3.7$, $\omega_{p}=0.75$ eV and $\gamma=0.118$ eV, which are possible values for heavily doped semiconductors, ZnO:Ga in our case~\cite{Kalusniak14}. For this scenario, according to \cref{eq:thickslab-nonretarded-limit}, the interaction still behaves at short separations as $\propto L^{-3}$ (black dashed line) but the impact of the thickness for $L\gtrsim d$ is more pronounced. 
For the non-ohmic conductor we obtain: $s^{(f)}_{p}=0.12$ for the full cylinder and $s^{(h)}_{p}=1.14 \times 10^{-10}$ for the hollow-core structure. According to these results, the perfect conductor behavior (black dash-dotted line) is only slowly recovered for $L\gg 10 R_{0}$ by the full cylinder comprised of the non-dissipative material. 
 Conversely, within the thin-wire limit, the hollow-core structure comprised of the non-ohmic conductor is well described by \cref{eq:superconductor-retarded-text} (black dashed curve). 
Considering an ohmic conductor, one has $s^{(f)}_{D}= 0.12$, while for the hollow-core structure, the condition for $s_{D}$ presented in \cref{eq:condition-conductor-retarded-max} is consistently satisfied within the thin-wire regime. 
Deviations from the preceding patterns, in accordance with \cref{eq:conductor-retarded} (black dashed curves), become notable at distances shorter than those observed for the good conductor, remaining tunable as a function of the shell thickness. 
}
\label{fig:casimir-polder-hollow-cylinder-conductors-distance-dependence}
\end{figure*}
%
%%%%%%%%%%%%%%%%%%%%%%%%%%%%%%%%%%%%%%%%%%%%%%
In \cref{fig:casimir-polder-hollow-cylinder-conductors-distance-dependence}, we
display the results of the numerical evaluation of the Casimir-Polder potential for a Rubidium atom interacting with an ohmic and a superconducting cylindrical shell. 
In order to highlight the interplay between the geometry and material parameters, as for the dielectric case, we compare the results for a full cylinder ($\chi=0$) and a cylindrical shell ($\chi=0.95$) with the same curvature ($R_0=2\lambdabar_a$) and thickness ($d=0.1\lambdabar_a$). 
Within the framework of our modeling in \cref{eq:permittivity-models}, we consider two sets of material parameters, one corresponding to a \textit{good conductor} (e.g. a noble metal such as gold or silver) and the other representing a \textit{poor conductor} (e.g. a doped semiconductor such as ZnO:Ga). The use of a poor conductor is motivated by the fact that the diffusion length $\lambda_{D}$ is enhanced for materials with higher resistivity, increasing the value for $s_{D}$ (see \cref{condition-conductor-retarded-full}) and thereby reducing the distance where, within the thin-wire limit, the trend described by \cref{eq:conductor-retarded} becomes visible.
As discussed below \cref{eq:permittivity-models}, to be able to directly compare the ohmic and the lossless conductors, we set $\gamma=0$ in the corresponding models describing the latter, keeping the same values for the remaining parameters.
In \cref{fig:casimir-polder-hollow-cylinder-conductors-distance-dependence} the different curves highlight how the interaction is affected by all of these aspects, identifying the limits of a description based on a perfect conductor. 

We first focus on the Casimir-Polder potential for the structures with material parameters characteristic of a good conductor. Analogously to \cref{fig:casimir-polder-hollow-core-cylinder-dielectric-distance-dependence}, for small distances the interaction recovers the behavior for a planar structure (see \cref{eq:thickslab-nonretarded-limit}): For the parameters considered here, both the ohmic and the non-dissipative conductor give effectively the same result, and the interaction cannot discern the nature of the conductor. 
Given the value of the radius, in this case the nonretarded limit is not compatible with the thin-wire limit, but we have independently checked that the corresponding asymptotic expressions derived in this section are consistent with our numerical evaluation. 
For larger distances, the effects of retardation and of the curvature become more pronounced.
Specifically, given the small value of $s_{D}$ and that $s_{p}\sim 1$ obtained for both $\chi$-values, within the thin-wire limit the impact of the material properties on the Casimir-Polder potential is negligible:
The interaction exhibits the same behavior as in the perfect conductor limit (dash-dotted line) presented in \cref{eq:perfect-conductor-retarded} and the shell thickness has a negligible effect.
Nonetheless, differently from the dielectric case, the transition from the power-law behavior of the slab limit to the logarithmic expression in \cref{eq:superconductor-retarded-text} is markedly visible for $L/R_{0}\gg s^{-1}_{p}$ in the log-log plot of Fig.~\ref{fig:casimir-polder-hollow-cylinder-conductors-distance-dependence}. 
For distances $L/R_{0}\gg s^{-1}_{D}$ the distinctive trend of \cref{eq:conductor-retarded} (black dashed curves), induced by the ohmic nature of the Drude model becomes apparent. 
Notably, in contrast to the nonretarded region, at large distances the interaction with an ohmic metal falls off faster than for the non-dissipative conductor.
In this regime, also the effects of the shell thickness are relevant. As discussed above, for a fixed position, they lead to a decrease in the interaction strength and also to a reduction of the minimal distance required to distinguish between the behavior of the ohmic and the non-dissipative conductor: For the parameters of a good conductor as considered in \cref{fig:casimir-polder-hollow-cylinder-conductors-distance-dependence}, the strength of the interaction for the corresponding value of $s_{D}$ is very similar.

When the parameters used for the structure are those of a poor conductor, the interplay between geometry and material properties has a more prominent effect. Once again, at short atom-cylinder separations the ohmic material and its superconducting variant behave almost identically. The potential follows the corresponding slab limit, and as for the dielectrics and differently from the good conductor, the impact of the shell thickness becomes more visible. At large separations the behavior is qualitatively similar to that of the good conductor but quantitatively, we observe the distinctive impact of the shell thickness. Indeed, given that for the full cylinder $s_{p}=0.12$, the perfect conductor limit of \cref{eq:perfect-conductor-retarded} is only reached at distances larger than ten times its radius.
For hollow-core cylinders, this value becomes even smaller, $s_{p}=1.14 \times 10^{-10}$, indicating that, within the thin-wire limit, the Casimir-Polder potential is well described by the expression in \cref{eq:superconductor-retarded-text}.
The impact of the ohmic behavior of the poor conductor on the Casimir-Polder potential is visible at a distance which is shorter than for the good conductor. In the case of the hollow-core cylinder, the condition connected with $s_{D}$ (see \cref{eq:condition-conductor-retarded-max}) is fulfilled for all distances within the thin-wire limit. For the full cylinder we have instead $s_{D}=0.12$, indicating in this case that the behavior described in \cref{eq:conductor-retarded} (black dashed curves) is already visible for distances larger than ten times its radius.

\section{Finite temperature corrections}
\label{sec:finite-temperature-interaction}

In order to complete our analysis of the Casimir-Polder interaction between an atom and a hollow-core cylinder, in this section we investigate how the previous results change when the system's temperature is no longer zero. 
As discussed at the beginning of \cref{sec:casimir-polder-interaction-in-multilayered-systems}, the main difference in the evaluation of the interaction lies in the fact that
\begin{equation}
\frac{\hbar}{2\pi}\int\limits_{0}^{\infty} \dd \xi  \to k_{B}T \primesum_{n=0}^{\infty}~,
\label{eq:zero-to-finite-temperature-replacement}
\end{equation}
i.e. the integral over imaginary frequencies is replaced by a discrete sum over the Matsubara frequencies (see \cref{eq:cp-interaction-finite-temperature}). 
For the specific configuration explored in this work, at finite temperature the only integral involved in the definition of the free energy is thus that over the wave vector $q$ appearing in the definition of the Green tensor (see \cref{eq:green-tensor-cylindrical-system}). 
Similar to the zero-temperature case, it is convenient to perform a change of variables as follows
\begin{equation}
\eta=\kappa L, \quad \frac{\xi_{n}}{c}L=n \frac{L}{\lambdabar_{T}}~,
\label{eq:variable-change-finite-temperature}
\end{equation}
where $\lambdabar_{T}=\hbar c/(2\pi k_{B}T)$($\sim 1.2$ $\mu$m at room temperature) is what we call the reduced Wien wavelength, since it corresponds to the characteristic length scale in Wien's displacement law divided by $2\pi$~\cite{Das02,Williams14}.
The integrand of the resulting expression for the free energy looks like the one in \cref{eq:cp-interaction-dimensionless-text}, divided by $\eta c/L$ and where we make the replacement $\zeta=(n L/\lambdabar_{T})/\eta$, i.e.
\begin{widetext}
\begin{align}
\begin{split}
\mathcal{F}
=
- 
\frac{k_{B} T}{\pi^2 } \frac{1}{L^{3}}
&\primesum_{n=0}^{\infty}
\int\limits_{n \frac{L}{\lambdabar_{T}}}^{\infty} \dd{\eta}
\primesum_{m=0}^{\infty} \;
 \frac{\alpha\left(\ii n\frac{c}{\lambdabar_{T}}\right)}{\epsilon_{0}}
  \frac{\mathcal{K}_{m}(\eta,s)}{\eta\sqrt{1-\frac{n^{2}\frac{L^{2}}{\lambdabar_{T}^{2}}}{\eta^{2}}}} 
  \Bigg\{~
     r_{m}^{\mathrm{NN}}
  \left( 
     1 +
     \left(1-\frac{n^{2} \frac{L^{2}}{\lambdabar_{T}^{2}}}{\eta^{2}}\right) \biggl[ \frac{m^2}{\eta^{2} [1+s]^2}
      +
     \mathbb{k}^{2}_m (\eta [1+s])
    \biggr]\right)
\\
  &{}-{} r_{m}^{\mathrm{MM}} \frac{n^{2}\frac{L^{2}}{\lambdabar_{T}^{2}}}{\eta^{2}}\left[ 
     \frac{m^2}{\eta^{2} [1+s]^2} 
     +
     \mathbb{k}^{2}_m(\eta [1+s])
 \right]
  + r_{m}^{\mathrm{MN}}
  4 
  \frac{n  \frac{L}{\lambdabar_{T}} }{\eta}
  \sqrt{1- \frac{n^{2} \frac{L^{2}}{\lambdabar_{T}^{2}}}{\eta^{2}}}
  \frac{ m }{\eta [1+s]} \mathbb{k}_m (\eta [1+s]) 
  \Bigg\}~.
\end{split}
\label{eq:cp-interaction-dimensionless-finite-temperature}
\end{align}
\end{widetext}
Again, the prime in the sums indicates that the $n=0$ and $m=0$ terms are to be taken with half their weight, respectively. 
As before, due to the behavior of $\mathcal{K}_{m}(\eta,s)$, the integrand is significantly different from zero only for $\eta\approx1$. However, the integration in the previous expression occurs only over $\eta\ge  n L/\lambdabar_{T}$, setting the constraint $n\lesssim \lambdabar_{T}/L$ on the relevant values of $n$. Consequently, the reduced Wien wavelength $\lambda_T$ defines two regimes. If $L\ll \lambdabar_{T}$, i.e. if the distance or the temperature are small enough, several terms of \cref{eq:cp-interaction-dimensionless-finite-temperature} fulfill the above constraint and contribute to the free energy in \cref{eq:cp-interaction-dimensionless-finite-temperature}. Within this range of distance, the resulting expression closely resembles the zero temperature limit in \cref{eq:cp-interaction-dimensionless-text}. This means that, despite $T\neq 0$, if $L\ll \lambdabar_{T}$, the corrections induced by the temperature are subleading and the system essentially behaves as if it were at zero temperature, recovering the same limiting expressions discussed in the analyses of the previous sections.
Conversely, for $L\gg \lambdabar_{T}$, i.e. for large distances or temperatures, only the term $n=0$ can fulfill the previous constraints, while the contribution of the remaining $n\ge 1$ terms becomes less relevant. 
The temperature effects then become significant and, in this thermal regime, \cref{eq:cp-interaction-dimensionless-finite-temperature} can be approximated by the $n=0$ term.

The value of the reduced Wien wavelength determines whether the thermal regime starts within the slab limit ($R_{0}/\lambdabar_{T}\gg 1$) or only overlaps with the thin-wire limit  ($R_{0}/\lambdabar_{T}\lesssim 1$). As above, we focus here only on the latter, where the effects of the surface curvature are more pronounced. We refer to the literature for the finite-temperature behavior of the Casimir-Polder free energy for an atom in front of a slab~\cite{Bostrom2000,Buhmann2012,Buhmann2012a}.
Considering only the $n=0$ term removes the contribution stemming from the MM and MN polarizations, leaving only that of $r_{m}^{\mathrm{NN}}$.
For reasons similar to those discussed at the beginning of \cref{subsec:thin-wire-T-zero}, in the thin-wire limit $s\ll 1$ only the orders $m=0,1$ are relevant (see \cref{eq:cp-interaction-finite-temperature-simplified}).
As for the zero-temperature limit, in the case of dielectrics, we require both $m$-values. In the thermal regime the Casimir-Polder interaction with the hollow-core cylinder then takes the asymptotic form
\begin{equation}
\label{eq:dielectric-finite-temperature}
\mathcal{F}\sim- \frac{9 k_{B}T}{512}  \frac{\alpha_{0}}{ \epsilon_{0}} \frac{R_{0}^{2}}{L^5}
\left[
\frac{1-\chi^{2}}{\Delta_{0}^{2}}
+ 6\frac{
\frac{1-\chi^{2}}{1-\frac{\chi^{2}}{\left[1+2 \Delta_{0}^2\right]^2}}
}{1+2 \Delta_{0}^2}\right]~.
\end{equation}
Once again, a non-vanishing $\Delta_0$ is central for arriving at the above expression. 
We note that the free energy looks similar to that observed in the zero-temperature retarded limit (see \cref{eq:cp-hollow-core-retarded-dielectric}), but with a power law identical to that of the nonretarded regime, i.e. it scales as $L^{-5}$ instead of as $L^{-6}$. 
Similar conclusions are therefore valid with respect to the impact of the thickness on the interaction strength, using the dimensional arguments discussed in the context of \cref{eq:dimensional-analysis}.
Due to the lack of terms other than those of the NN-polarization, the proportion between the contribution stemming from the orders $m=0$ and $m=1$ is altered with respect to \cref{eq:cp-hollow-core-retarded-dielectric}.
%%%%%%%%%%%%%%%%%%%%%%%%%%
\begin{figure}
\centering
\includegraphics[width=\columnwidth]{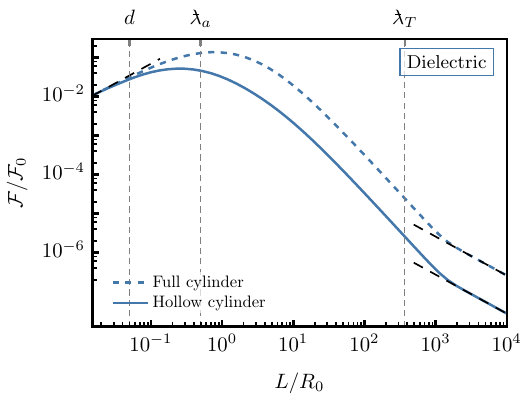}
\caption{Casimir-Polder free energy for a Rubidium atom interacting with a hollow-core silica fiber. 
The geometrical and material parameters are the same as in \cref{fig:casimir-polder-hollow-core-cylinder-dielectric-distance-dependence-mdl-comparison}. The temperature of the system is $4$ K. This implies $R_{0}/\lambdabar_{T}\ll 1$, indicating that thermal effects are visible only within the thin-wire limit. 
At distances shorter than the reduced Wien wavelength the behavior is very close to that obtained in \cref{fig:casimir-polder-hollow-core-cylinder-dielectric-distance-dependence} in the zero temperature limit. 
 Within the thin-wire limit for $R_{0}\ll L\ll \lambdabar_{T}$, the potential starts to decrease as $\propto L^{-6}$ according to \cref{eq:cp-hollow-core-retarded-dielectric} to then follow for $L\gg \lambdabar_{T}$ the $\propto L^{-5}$ behavior predicted in \cref{eq:dielectric-finite-temperature} (black dashed curves).}
\label{fig:dielectric-distance-dependence-finite-temperature}
\end{figure}
\begin{figure}[ht!]
  \subfloat[]{
    \includegraphics[width=\columnwidth]{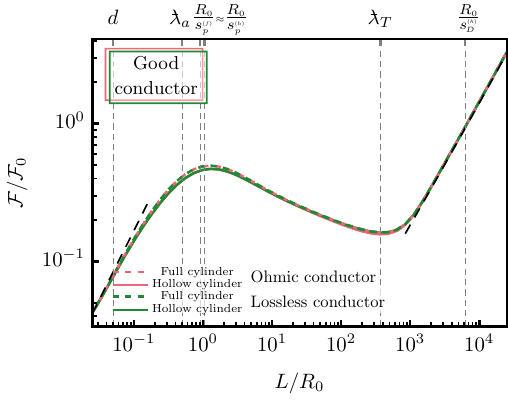}
  }
  \hspace{0.1\columnwidth}
  \subfloat[]{
    \includegraphics[width=\columnwidth]{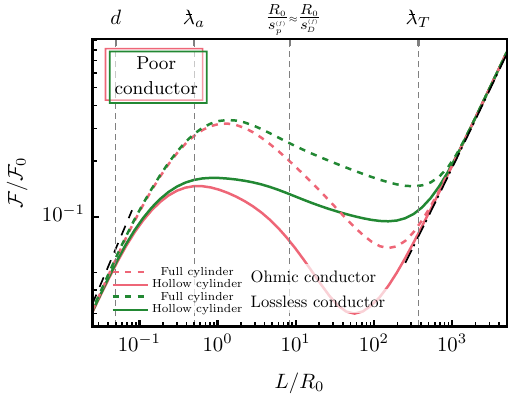}
  }
\caption{Finite-temperature Casimir-Polder free energy for a Rubidium atom interacting with a cylindrical shell comprised of a) a good and b) a poor conductor. The geometry and material parameters are the same as in \cref{fig:casimir-polder-hollow-cylinder-conductors-distance-dependence}. The interaction is depicted as a function of the atom-structure separation. The temperature of the system is $4$ K. 
For $L\ll \lambdabar_{T}$, despite a nonzero temperature, the free energy behaves as if $T=0$.  If sufficiently distant from the cylindrical structure ($L\gg \lambdabar_{T}$), the interaction enters the thermal regime, behaving in agreement with \cref{eq:conductor-finite-temperature} (black dash-dotted lines). In this region, all the information about the thickness of the structure and of the material parameters is lost. For the parameters considered here, at a temperature of $\SI{4}{\kelvin}$, the ohmic and non-ohmic good conductors give rise to effectively the same Casimir-Polder potential, independently from the specific geometry of the cylindrical structure. Differently, for the poor conductor parameters, the differences between the ohmic and the non-ohmic material are still visible despite the finite temperature.}
\label{fig:cp-distance-dependence-conductors-finite-temperature}
\end{figure}
%%%%%%%%%%%%%%%%%%%%%%%%%%
As in the zero-temperature case, \cref{eq:dielectric-finite-temperature} is not valid for conductors, and would indeed diverge since for this case we have $\Delta_0 \to 0$.
Similarly to the case of a conducting cylindrical shell at zero temperature, an inspection of \cref{eq:cp-interaction-dimensionless-finite-temperature} reveals that the leading contribution to the Casimir-Polder free energy arises in this case only from the $m=0$ term. 
Here, however, since we are setting $\zeta=0$ from the beginning, $\Delta$ always identically vanishes independent of the type of conductor. 
This leads to a situation which is to some extent similar to that discussed in the previous section: A vanishing $\Delta$ removes all information about the physical parameters that describe the material comprising the cylindrical structure, making it effectively equivalent to a perfect conductor. 
We have indeed that in the thermal regime, no matter which of the material or geometrical parameters we use, the Casimir-Polder free energy for a conductor behaves as (c.p.\ also~\cite{Barash1989})
\begin{align}
\mathcal{F}
&\sim
 - 
\frac{k_{B}T}{32}  \frac{\frac{\alpha_{0}}{\epsilon_{0}}}{L^3\ln\left(\frac{\frac{L}{R_{0}}}{\eta_{[2]} \tilde{\gamma}_{\mathrm{E}}} \right)}
~.
\label{eq:conductor-finite-temperature}
\end{align}
\Cref{eq:dielectric-finite-temperature} indicates that the thermal regime washes out the differences between ohmic and non-ohmic conductor highlighted in the zero temperature case: Since they both effectively behave as a perfect conductor, the influence of the shell thickness on the final outcome is erased. 
The interaction scales linearly with $T$ and acquires a logarithmic behavior as a function of $L$ which is similar to that observed at zero temperature in the nonretarded limit (see \cref{eq:superconductor-nonretarded-1}). 
Notice that, given the general expression of the Casimir-Polder free energy at finite temperature in \cref{eq:cp-interaction-finite-temperature}, most of the previous reasoning usually remains valid for other geometries~\cite{Buhmann2012a}. Unlike a dielectric, which permits field penetration, a conductor screens its internal structure from the interaction when the atom-object separation significantly exceeds $\lambdabar_{T}$. The interaction is then only affected by the geometry of the interface with vacuum, resulting in a logarithmic scaling given in Eq. (42) for a cylindrical structure (power-law behaviors are commonly found in other geometries~\cite{Buhmann2012a}).

In \cref{fig:dielectric-distance-dependence-finite-temperature,fig:cp-distance-dependence-conductors-finite-temperature} we present the Casimir-Polder free energy for both dielectric and conducting materials at finite temperature. The geometrical and the material parameters are the same of Figs. \ref{fig:casimir-polder-hollow-core-cylinder-dielectric-distance-dependence-mdl-comparison} and \ref{fig:casimir-polder-hollow-cylinder-conductors-distance-dependence}, respectively. The temperature is 4 K, corresponding to a reduced Wien wavelength ($\approx 91$ $\mu$m) which is about four hundred times larger than the value for the external radius of the cylinder considered here.
In the case of dielectrics, as expected the behavior of the interaction is close to the zero temperature limit for $L\ll \lambdabar_{T}$ and, according to \cref{eq:dielectric-finite-temperature}, the impact of the temperature is visible in a change of the power law from $L^{-6}$ to $L^{-5}$ occurring for $L\gg \lambdabar_{T}$.
For the parameters used to describe a good conductor, our numerical evaluation confirms that the temperature almost completely erases all the differences between the ohmic and the non-ohmic variant, independent of whether the hollow-core or the full cylinder is considered. Using the value for $s_{D}$ corresponding to the cylindrical shell, we can estimate that the difference between the ohmic and the non-ohmic conductor would appear for temperatures well below $\approx 0.24$ K, supporting therefore the behavior pictured in \cref{fig:cp-distance-dependence-conductors-finite-temperature}. The curves also display the asymptotic expression given in \cref{eq:conductor-finite-temperature} valid for $L\gg \lambdabar_{T}$.
The behavior changes in the case of a poor conductor: Our numerical evaluation highlights that for $T=\SI{4}{\kelvin}$, differences between the ohmic and non-ohmic material are still discernible for both cylindrical structures. This is in line with the estimates provided by $s_{p}$ and $s_{D}$ for the parameters of a poor conductor. In particular, $R_{0}/s_{D}<\lambdabar_{T}$, i.e. the distance at which characteristic features induced by the ohmic material become apparent is shorter than the characteristic length from which the thermal effect becomes dominant, thereby eliminating the distinction between the conductors. While in the case of the hollow-core cylinder the effect should be visible even at room temperature or higher, for the full cylinder our calculation points to temperatures significantly smaller than about $\SI{180}{\kelvin}$.
However, also in the case of poor conductors, for $L\gg \lambdabar_{T}$ the Casimir-Polder potential follows the behavior described in \cref{eq:conductor-finite-temperature}, losing all information about the shell thickness and the material properties.

\section{Conclusions and discussion}
\label{sec:conclusion}

In this work we have provided an in-depth investigation of the Casimir-Polder interaction between an atom and hollow-core cylindrical structures.
This type of geometry often appears in modern experiments~\cite{Vetsch10,Solano17a,Le-Kien18,Langbecker17,LeKien2022,Fermani07,Gierling11,Schneeweiss12}, but it has been scarcely addressed in the literature and with limited detail.
Our analyses have highlighted that the Casimir-Polder potential is modified in a very distinctive way by the interplay between geometry and material properties.

Specifically, we have considered a cylindrical shell comprised of spatially local and isotropic dielectrics and conductors. 
Compared to the more frequently investigated planar structures, infinitely extended in two directions, this geometry features a change of dimensionality due to an additional spatial dimension being constrained as prescribed by the related curvature radius $R_{0}$. 
Moreover, it shows an additional geometrical signature associated with confinement induced by the shell thickness $d$. The materials addressed in our investigation have been conveniently characterized in terms of their general low-frequency behavior and how the electromagnetic field penetrates the structure. 
To this end, we have defined the relative penetration depth $\Delta$ (cf. \cref{eq:relative-penetration-depth}), which can exhibit significant variations for dielectrics and distinct types of conductors. 
We have connected this aspect with the underlying physical phenomena via the typical length scales that characterize the specific material. 
The majority of our conclusions are therefore not tied to a specific model for the atomic polarizability or the dielectric permittivity and lend themselves to possible generalizations.  

We have analyzed the expressions that describe the Casimir-Polder free energy both numerically and analytically, for different materials and geometrical parameters as a function of the separation $L$ between the atom and cylindrical structure. 
For our numerical investigations, in \cref{sec:numerical-evaluation} we have developed an approach based on a carefully crafted multi-variate Gaussian quadrature scheme~\cite{Engblom2006,Monien2010} which efficiently and flexibly provides converging results over a large range of separations between the atom and the hollow-core cylinder, despite the very distinct length scales characterizing the system.
Analytically, we distinguished between the slab and the thin-wire limits: They identify two regimes for which $R_{0}/L\gg 1$ and $R_{0}/L\ll 1$, respectively. In the slab limit the hollow-core cylinder effectively appears as a dielectric slab with thickness $d$, recovering the results known in the literature~\cite{Intravaia2011,Buhmann2012,Buhmann2012a}. 
Therefore, our attention has been mostly focused on the second regime, where the cylindrical structure effectively behaves as a 1D infinitely extended wire with vanishing radius.
The corresponding distance range was examined in the nonretarded and in the retarded regions, defined in terms of the wavelength of the characteristic atomic electromagnetic transition. 
In particular, we have elucidated when retardation effects can and when they cannot be neglected.

Using this combination of numerical and analytical tools we were able to show that for a dielectric cylindrical shell, the Casimir-Polder interaction features a power-law behavior with characteristic exponents that depend on the different regions: The scaling of the interaction varies from $\propto L^{-3}$ at very short distances within the slab limit to $\propto L^{-5}$ and $\propto L^{-6}$ in the nonretarded and the retarded region of the thin-wire limit, respectively, to become again $\propto L^{-5}$ as soon as the thermal effects become relevant.
Despite the strong dependence on the detailed form of the interaction on the properties of the system [see for example Eqs.~\eqref{eq:cp-hollow-core-nonretarded}, \cref{eq:cp-hollow-core-retarded-dielectric} and \cref{eq:dielectric-finite-temperature}], it is worth to mention that the change of exponent in the power law can be understood in terms of dimensional analyses (see \cref{eq:dimensional-analysis}): The increase in relevance of an additional length scale, induced by a geometric constraint, i.e. curvature and thickness, must be dimensionally compensated for by a change in the power-law exponent of the distance dependence. Specifically, within the thin-wire limit, the Casimir-Polder free energy scales with the cross-sectional area of the cylindrical shell, prominently showcasing how the shell thickness can be used to control the interaction strength even in the limit when the cylinder looks like a wire.

The situation is very different for conductors. Due to the presence of conduction electrons, the electromagnetic radiation is more efficiently screened from the inside of the cylindrical structure. This manifests as a drastic reduction of the penetration depth and in the appearance of new length scales connected with the screening as well as the diffusive dynamics of the radiation in the material (see the end of \cref{sec:casimir-polder-interaction-in-multilayered-systems}). 
In turn, this leads to a modification of the (quantum) fluctuations of the electromagnetic field, consequently affecting the behavior of the Casimir-Polder interaction. Specifically, the interplay between material properties and geometry exhibit rather unusual characteristics relative to those obtained for dielectrics and for planar structures. In fact, while for a planar structure the interaction still features a conventional power-law behavior $\propto L^{-3}$ in the nonretarded limit and $\propto L^{-4}$ when retardation becomes relevant~\cite{Casimir1948}, we have demonstrated that as soon as the curvature and the thickness of the cylindrical shell cannot be disregarded, the Casimir-Polder potential features different logarithmic forms, depending on the combination of the different physical mechanisms at work in the system. In particular, in the thin-wire limit the Casimir-Polder interaction is able to prominently distinguish whether a conductor is ohmic or not. Ohmic behavior gives rise to a characteristic square-root behavior for the penetration depth at low frequencies (see \cref{eq:relative-penetration-depth-materials-zero-frequency}). This aspect is related to the existence of a nonvanishing zero-frequency resistivity of the material.
If this behavior is suppressed, by considering for example a model for a non-ohmic material with zero DC-resistivity, the values of the free energy for ohmic and non-ohmic conductors start to differ for some specific regions of the atom-cylinder separation. These regions are roughly bounded from below by the slab limit and from above by the thermal regime, which is the region where the impact of the temperature can no longer be neglected. 
In our analyses, we have focused on only one type of non-ohmic behavior, which corresponds to a non-dissipative material. Dissipation is central in a longstanding controversy surrounding the Casimir interaction~\cite{Klimchitskaya22c} and our results can provide additional information about this issue.
For the same cylinder radius $R_{0}$, within the thermal region the interaction becomes insensitive to all differences between material parameters of the conducting material: It is only relevant that the penetration length vanishes at zero frequency, which is valid for any kind of conductor, including idealized ones. For distances larger than the reduced Wien wavelength the interaction behaves as if the cylindrical structure was comprised of a perfect conductor, thereby effectively losing all information about the thickness of the cylindrical shell.

 In our work, we have been able to formally specify the region of distance where the differences in the free energy induced by the ohmic and the non-ohmic material occur. In particular, through the definition of the parameter $s_{D}$, we have shown how the lower boundary of this region connects with all other (geometrical and physical) length scales of the system. 
 One of the main outcomes of our analyses is also that the shell thickness can be effectively used to not only tailor the interaction strength but to also shift this boundary to smaller atom-surface separations, suggesting this geometrical feature as a versatile tool for controlling the interaction and for making the previous region experimentally accessible.
From a technical perspective, our findings can also be beneficial for the analysis of near-field heat transfer~\cite{Asheichyk22,Wang24a,Wang25a}, electron-energy-loss spectroscopy (EELS)~\cite{Zhao24b,Rodriguez-Echarri25,Egerton09}, and other areas of physics~\cite{Rekdal04,Ruppin01,Dedkov21,Reiche22,Wegner23}, where the knowledge of the electromagnetic Green tensor for cylindrical structures is pertinent.

The present work serves as a foundation for more comprehensive investigations. While the current framework is broadly applicable, we can achieve further refinements by incorporating more sophisticated material models that extend beyond a local description of the dielectric function.
Beyond the numerous intriguing phenomena in photonics~\cite{Yan12,Monticone25}, previous studies have demonstrated that nonlocality (spatial dispersion) introduces additional length scales into the system and can significantly impact the physics of Casimir interactions~\cite{Esquivel03,Contreras-Reyes05,Svetovoy08a,Reiche20,Kristensen23,Zhang25} and fluctuation-induced phenomena in general~\cite{Chapuis08,Haakh12,Intravaia15a,Schmidt16,Reiche17,Reiche19}. In particular, in nanoscopic systems, the resulting confinement of conduction electrons coupled to their dynamics within a conductor can lead to a difference in the material’s optical response and related effects within the near-field region of the conducting body (typically within a few tens of nanometers from the material interface).
Although this may necessitate a revision of the results within the slab limit, we expect that the remaining outcomes, including the ability of our setup to distinguish between the types of conductor, should be preserved to a significant extent. However, due to the intricate interplay of various physical aspects characterizing the system, such as the quasi-one-dimensional constraint in the thin-wire limit and a possibly very thin shell, a detailed analysis is required to confirm this behavior.
On the side of the particle, we can further consider microscopic objects with anisotropic polarizabilities. 
In this case, the nontrivial tensorial structure of the latter requires the separate consideration of individual Green tensor components in accordance with Eq. (1) (see also \cite{Eberlein2007,Eberlein2009}), which can influence the interaction's scaling behavior and even result in novel effects.
Specifically, previous works with perfectly conducting cylinders~\cite{Milton12,Alves23} 
demonstrated that in this case the Casimir-Polder interaction can exhibit what can be called ``directional repulsion''. This means that the force can change from attractive to repulsive, depending on the direction considered. Our approach offers the flexibility to investigate how directional repulsion is modified not only when more realistic material models are taken into account but also when the shell thickness is introduced as an additional tuning parameter. Alternatively, one could consider atoms
in excited states rather than at equilibrium with their surroundings, a configuration also known to produce repulsive interactions~\cite{Buhmann2012a,Le-Kien22}.
The results presented here can serve as a benchmark for independently verifying and evaluating the accuracy of recently developed full numerical schemes~\cite{Reid09,Kristensen23,Emig23,Bimonte23,Bimonte25b} for a nontrivial geometry. Conversely, the flexibility of these numerical schemes can be utilized to assess the range of validity of our result.
As an example, an atom at a distance comparable
to or significantly larger than the fiber’s length can no longer be described within the infinitely extended wire approximation. 
Instead, we anticipate a transition to a particle-like regime with the interaction progressively recovering a power-law behavior.

Specific to the Casimir-Polder interaction, since it induces a shift in the level structure of the atom, a measurement of the corresponding effects can allow to identify the nature of the material comprising the structure~\cite{Fiedler20a}. 
Furthermore, the setup we consider is employed in numerous experiments~\cite{Reichel99,Fortagh07,Henkel09,Keil16,Vetsch10,Solano17a,Le-Kien18,Langbecker17,LeKien2022,Fermani07,Gierling11,Schneeweiss12,Lodahl17,Meng18,Hummer21,Lechner23}, which already provide the hardware for future quantum sensors. 
In an era where spectroscopic and in general frequency-related investigations of atomic systems provide some of the most precise measurements, the outcomes presented in our manuscript, in conjunction with possible parameter optimizations, will open new pathways for a better understanding of the Casimir-Polder interaction. 
Given that this quantum mechanical phenomenon is situated at the crossroads of different areas of science, this will also offer new opportunities for both fundamental investigations and 
quantum technologies. 

\acknowledgements
F.I. acknowledges BERLIN QUANTUM, an initiative endowed by the Innovation Promotion Fund of the city of Berlin, for financial support.

%%%%%%%%%%%%%%%%%%
%%% Appendices %%%
%%%%%%%%%%%%%%%%%%

\begin{appendices}
\crefalias{section}{appendix}

\appendix
\section{The scattered Green tensor for cylindrical geometries}
\label{app:green-tensor}

In this work, we define the electric field Green tensor $\tensor{G}_{\mathrm{t}}(\vec{r}, \vec{r}', \omega)$ in a spatially local nonmagnetic medium described by the  relative permittivity $\epsilon(\omega)$ and relative permeability $\mu(\omega)=1$ as the solution to the vector wave equation 
\begin{equation}
\label{eq:green-tensor-vector-wave-eq-definition} 
\nabla \times \nabla \times \tensor{G}_{\mathrm{t}}(\vec{r}, \vec{r}', \omega)
- 
\frac{\omega^2}{c^2} \epsilon(\omega) \tensor{G}_{\mathrm{t}} (\vec{r}, \vec{r}', \omega)
=
\frac{\omega^2}{\epsilon_{0} c^2}
\delta(\vec{r} - \vec{r}') \underline{\mathds{1}}
,
\end{equation}
where $\epsilon_{0}$ is the vacuum permittivity and $c$ is the speed of light. 
Physically, it describes the electric field at position $\vec{r}$ generated by an oscillating point dipole source located at position $\vec{r}'$. 
We note that, differently from the convention sometimes used in the literature \cite{Buhmann2012,Buhmann2012a}, the right hand side of \cref{eq:green-tensor-vector-wave-eq-definition} includes a prefactor in front of the Dirac delta function. 
When comparing to references using this alternative definition, one therefore has to multiply the expressions presented in this appendix with a factor of $\epsilon_{0} c^2 / \omega^2$.

For a typical scattering problem, space is partitioned into several regions, defining the boundary between the scattering objects and the environment.
Equation \eqref{eq:green-tensor-vector-wave-eq-definition} then holds individually for each region, with the relative permittivity understood to be defined piecewise accordingly.
In general, the Green tensor can be split into a free space part, $\tensor{G}_{0}$, and a scattered part, $\tensor{G}$, such that 
\begin{equation}
\tensor{G}_{\mathrm{t}}(\vec{r},\vec{r}',\omega) = \tensor{G}_{0}(\vec{r},\vec{r}',\omega) + \tensor{G}(\vec{r},\vec{r}',\omega)~.
\end{equation}
To calculate the Casimir-Polder interaction, we must evaluate the Green tensor in the coincidence limit, i.e., $\vec{r} = \vec{r}'$, in which case only the scattered part depends on position. 
Consequently, also in relation to the force that can be derived from the potential, only $\tensor{G}$ is usually associated with the Casimir-Polder free energy since it directly defines the spatial dependence of the interaction. 
In addition, this removes the formal divergence usually occurring for $\tensor{G}_{0}$ in the coincidence limit~\cite{Intravaia2011,Buhmann2012,Buhmann2012a}.

Let us consider the case of a multilayered cylindrical structure having its axis of rotational symmetry coinciding with the $z$-axis of our reference frame.
Due to the structure's translational and rotational symmetry, we can always identify a generic position $\vec{r}$ by knowing the radius of the outer-most layer, $R_{0}$, and the radial distance of the atom to the cylinder's surface, $L$.
It can be shown that in dyadic notation, we have (c.p.~\cite{Tai1994})
\begin{widetext}
\begin{equation}
\label{eq:green-tensor-cylindrical-system} 
\begin{aligned}
 \tensor{G}(L, \ii\xi)
=
\frac{1}{\pi^{2}\epsilon_{0}}
\int\limits_{0}^{\infty} \dd{q} \,
\primesum_{m=0}^\infty
     &q^2 \frac{I_m(\kappa s L)}{K_m(\kappa s L)}  K^2_m(\kappa[1+s]L)  
\\     
    \times  \Bigg\{
   &r_{m}^{\mathrm{NN}}(\ii\xi, q)\; 
    \bigg[
      \mathbb{k}^{2}_m(\kappa[1+s]L) \vec{\hat{e}}_{\rho}\vec{\hat{e}}_{\rho}
      +
       \left(\frac{m}{\kappa[1+s]L}\right)^2 \vec{\hat{e}}_\phi\vec{\hat{e}}_\phi
      +
       \frac{\kappa^2}{q^2}\vec{\hat{e}}_z\vec{\hat{e}}_z
    \bigg]
\\
    - &r_{m}^{\mathrm{MM}}(\ii\xi, q) \;
     \frac{\xi^2}{c^{2}q^{2}} \left[
      \left(\frac{m}{\kappa[1+s]L}\right)^2 \vec{\hat{e}}_{\rho}\vec{\hat{e}}_{\rho}
      +
      \mathbb{k}^{2}_m(\kappa[1+s]L) \vec{\hat{e}}_\phi\vec{\hat{e}}_\phi
    \right]   
\\
    + 2 &r_{m}^{\mathrm{MN}}(\ii\xi, q) 
   \frac{\xi}{c q} 
     \left(\frac{m}{\kappa[1+s]L}\right)
  \mathbb{k}_m(\kappa[1+s]L) 
    \left[ \vec{\hat{e}}_{\rho}\vec{\hat{e}}_{\rho} + \vec{\hat{e}}_\phi\vec{\hat{e}}_\phi \right]
  \Bigg\}~,
\end{aligned}
\end{equation}
\end{widetext}
with the cylindrical unit vectors $\{\vec{\hat{e}}_{\rho},\vec{\hat{e}}_{\phi},\vec{\hat{e}}_{z}\}$. As mentioned in \cref{subsec:em-field-scattering-cylinder}, the functions $r_{m}^{ij}$, ${i,j\in\{\mathrm{N},\mathrm{M}\}}$ are the scattering coefficients characterizing the electromagnetic scattering behavior of the multilayered cylinder. 

For the purpose of our calculation, all expressions are evaluated at imaginary frequencies $\omega\to\ii\xi$.
Since the Green tensor represents a susceptibility, for imaginary frequencies all its elements are given by real functions.
In the above expression, $s=R_{0}/L$, $q$ is the modulus of the wave vector component along $\vec{\hat{e}}_{z}$ and $\kappa = \sqrt{q^2 +\xi^2/c^{2}}$ is related to the wave vector component perpendicular to the surface. 
In our analysis, this Green tensor is introduced into \cref{eq:cp-interaction-zero-temperature}, which gives the Casimir-Polder free energy. Given that our description is focusing on isotropic polarizabilities, the trace in \cref{eq:cp-interaction-zero-temperature} effectively sums the diagonal terms of $\tensor{G}$. 

To obtain \cref{eq:cp-interaction-dimensionless-text} we have performed the following change of variable:
\begin{subequations}
\label{change-of-variables}
\begin{equation}
\frac{\xi}{c}L =\zeta \eta\quad \text{and}\quad \eta =\kappa L ~.
\end{equation} 
Then, the double integration over $\xi$ and $q$ that appears in the expression for the Casimir-Polder free energy after inserting the Green tensor in \cref{eq:green-tensor-cylindrical-system} into \cref{eq:cp-interaction-zero-temperature}, can be conveniently rewritten using the following identities:
\begin{equation}
\begin{aligned}
&\int\limits_{0}^{\infty}\dd{\xi} \int\limits_{0}^{\infty}\dd{q}\; q^{2}f(\xi,q)
\\
&=\frac{c}{L^{4}}\int\limits_{0}^{\infty}\dd{\eta}  \int\limits_{0}^{1}\dd{\zeta}\; \frac{\eta^{3}(1-\zeta^{2})}{\sqrt{1-\zeta^{2}}}  \tilde{f}(\zeta,\eta)~,
\end{aligned}
\end{equation}
where $f$ denotes the integrand before the change of variables  and $\tilde{f}$ is the corresponding expression in the new variables. In accordance with this change of variables 
we have defined
\begin{equation}
\eta_{\epsilon}=\kappa_{\epsilon}L=\eta\sqrt{1+\frac{\zeta^{2}}{\Delta^{2}\left(\ii\zeta\frac{c}{R_{0}}\eta s\right)}}
\end{equation}
\end{subequations}
where $\Delta(\ii \xi)$ is the function introduced  in \cref{eq:relative-penetration-depth}. In fact, $\Delta(\ii \xi)$ is related to the material's relative permittivity and more precisely to
the penetration depth of the electromagnetic radiation.

In \cref{eq:green-tensor-cylindrical-system} -- and consequently in \cref{eq:cp-interaction-dimensionless-text} -- the 
quantity $m \in \mathbb{N}_{0}$ is the order of the modified Bessel functions of the first and second kind, $I_m$ and $K_m$ and of their corresponding logarithmic derivatives, $\mathbb{i}_m$ and $\mathbb{k}_m$. 
For fixed order $m$ and $x\to\infty$, these functions have the following asymptotic properties~\cite{Abramowitz71}:
\begin{subequations}
\label{eq:asymptotic-expressions}
\begin{align}
I_m(x)&\sim
\begin{cases}
\frac{1}{m!}\left(\frac{x}{2}\right)^{m}& x\to 0
\\
\\
\frac{1}{\sqrt{2\pi x}}e^{x}& x\to \infty~,
\end{cases}
\\
\nonumber
\\
K_m(x)&\sim
\begin{cases}
-\ln\left(x \tilde{\gamma}_{\mathrm{E}}\right) & x\to 0\quad  \text{for } m=0
\\
\\
\frac{(m-1)!}{2}\left(\frac{x}{2}\right)^{-m} & x\to 0
\\
\\
\sqrt{\frac{\pi}{2x}}e^{-x}& x\to \infty
\end{cases}
\label{eq:modified-bessel-argument-asymptotics}
\end{align}
where $\tilde{\gamma}_{\mathrm{E}}=e^{\gamma_{\mathrm{E}}}/2$ and $\gamma_{\mathrm{E}}$ is the Euler–Mascheroni constant.
For the corresponding logarithmic derivatives we thus obtain
\begin{align}
\mathbb{i}_m(x)&\equiv\frac{I'_m(x)}{I_m(x)} 
\sim
\begin{cases}
\frac{x}{2} & x\to 0 \quad \text{for } m=0
\\
\frac{m}{x} & x\to 0
\\
1 & x\to \infty~,
\end{cases}
\\
\nonumber
\\
\mathbb{k}_m(x)&\equiv\frac{K'_m(x)}{K_m(x)}
\sim
\begin{cases}
\frac{1}{x\ln\left(x \tilde{\gamma}_{\mathrm{E}}\right)}& x\to 0 \quad  \text{for } m=0
\\
- \frac{m}{x} & x\to 0
\\
-1 & x\to \infty~.
\end{cases}
\label{eq:modified-bessel-logarithmic-derivatives-asymptotics}
\end{align}
\end{subequations}
Moreover, $I_m$ and $K_m$ are both strictly positive for a positive argument and order. In general, for orders $\nu \in \mathbb{R}$ they fulfill the following bounds~\cite{Segura21,Baricz10}
\begin{subequations}
\label{eq:bessel-bounds}
\begin{equation}
I_{0}(x)K_{0}(x)<\frac{1+8x}{16 x^{2}}~,
\end{equation}
\begin{equation}
I_{\nu}(x)K_{\nu}(x)<\frac{1}{2\sqrt{\left(\nu-\frac{1}{2}\right)^{2}+x^{2}}}, \quad (\nu>1/2)~,
\end{equation}
\begin{equation}
0<\frac{K_{\nu}(y)}{K_{\nu}(x)}< \left(\frac{x}{y}\right)^{\nu}, \quad (0<x<y,\; \nu>-1/2)~,
\end{equation}
\begin{equation}
\left(\frac{y}{x}\right)^{\nu}<\frac{I_{\nu}(y)}{I_{\nu}(x)}< e^{y-x}\left(\frac{y}{x}\right)^{\nu}, \quad (0<x<y,\; \nu>-1/2)~.
\end{equation}
For their logarithmic derivatives this leads to \cite{Amos74,Yang17a}
\begin{equation}
0<\mathbb{i}_{\nu}(x)\le \sqrt{1+\frac{\nu^{2}}{x^{2}}}~,
\end{equation}
\begin{equation}
-\sqrt{\frac{\nu}{\nu-1}+\frac{\nu^{2}}{x^{2}}}\le \mathbb{k}_{\nu}(x)\le -\sqrt{1+\frac{\nu^{2}}{x^{2}}} \quad (\nu>1)~.
\end{equation}
\end{subequations}
The right hand side of both the above inequalities is also valid for an arbitrary $\nu \in \mathbb{R}$. 
\\

\subsection{Scattering coefficients for a cylindrical shell}
For the specific case of a cylindrical shell, the expressions for the scattering coefficients are provided in Eqs.~\eqref{eq:scattering-coefficients-hollow-cylinder}.
For convenience we reproduce them here in terms of the original variables $\xi$, $\kappa$ and $\kappa_{\epsilon}$:
\begin{subequations}
\begin{align}
r_{m}^{\mathrm{NN}}
&=\frac{\frac{\epsilon \Phi_{m}^{(\epsilon)} -\frac{\kappa_{\epsilon}}{\kappa} \frac{\mathbb{i}_m(\kappa R_0)}{\mathbb{i}_m(\kappa_{\epsilon} R_0)}}{\Pi_{m}^{21}}
+
 \frac{ \left(\frac{q \frac{\xi}{c}}{\kappa^{2}\Delta^{2}}\frac{\frac{m}{\kappa_{\epsilon} R_0}}{\mathbb{i}_m(\kappa_{\epsilon} R_0)} \right)^{2}
}{\Phi_{m}^{(1)}-\frac{\kappa_{\epsilon}}{\kappa}\frac{\mathbb{k}_m(\kappa R_0)}{\mathbb{i}_m(\kappa_{\epsilon} R_0)} }
}
{\frac{ \epsilon \Phi_{m}^{(\epsilon)} -\frac{\kappa_{\epsilon}}{\kappa} \frac{\mathbb{k}_m(\kappa R_0)}{\mathbb{i}_m(\kappa_{\epsilon} R_0)}}{\Pi_{m}^{22}}
+
\frac{\left(\frac{q \frac{\xi}{c}}{\kappa^{2}\Delta^{2}}\frac{\frac{m}{\kappa_{\epsilon} R_0}}{\mathbb{i}_m(\kappa_{\epsilon} R_0)}  \right)^{2}
}{\Phi_{m}^{(1)}-\frac{\kappa_{\epsilon}}{\kappa}\frac{\mathbb{k}_m(\kappa R_0)}{\mathbb{i}_m(\kappa_{\epsilon} R_0)}}
}
~,
\end{align}

\begin{align}
r_{m}^{\mathrm{MM}}
&=\frac{
\frac{\Phi_{m}^{(1)}-\frac{\kappa_{\epsilon}}{\kappa} \frac{\mathbb{i}_m(\kappa R_0)}{\mathbb{i}_m(\kappa_{\epsilon} R_0) }}{\Pi_{m}^{12}}+  
\frac{\left( \frac{q \frac{\xi}{c}}{\kappa^{2}\Delta^{2}}\frac{\frac{m}{\kappa_{\epsilon} R_0}}{\mathbb{i}_m(\kappa_{\epsilon} R_0)} \right)^{2}}
{\epsilon\Phi_{m}^{(\epsilon)} -\frac{\kappa_{\epsilon}}{\kappa} \frac{\mathbb{k}_m(\kappa R_0)}{\mathbb{i}_m(\kappa_{\epsilon} R_0)}}}
{
\frac{\Phi_{m}^{(1)}-\frac{\kappa_{\epsilon}}{\kappa}\frac{\mathbb{k}_m(\kappa R_0)}{\mathbb{i}_m(\kappa_{\epsilon} R_0)}}{\Pi_{m}^{22}} +  
\frac{\left( \frac{q \frac{\xi}{c}}{\kappa^{2}\Delta^{2}}\frac{\frac{m}{\kappa_{\epsilon} R_0}}{\mathbb{i}_m(\kappa_{\epsilon} R_0)} \right)^{2}}
{ \epsilon \Phi_{m}^{(\epsilon)}-\frac{\kappa_{\epsilon}}{\kappa} \frac{\mathbb{k}_m(\kappa R_0)}{\mathbb{i}_m(\kappa_{\epsilon} R_0) }}}
~,
\end{align}

\begin{align}
r_{m}^{\mathrm{MN}}
&=-\frac{\frac{1}{\Pi_{m}^{0}} 
\frac{\left(\frac{q \frac{\xi}{c}}{\kappa^{2}\Delta^{2}} \frac{\frac{m}{\kappa_{\epsilon} R_0}}{\mathbb{i}_m(\kappa_{\epsilon} R_0)} \right)\left[
\frac{\kappa_{\epsilon}}{\kappa}\frac{\mathbb{i}_{m}(\kappa R_{0})}{\mathbb{i}_m(\kappa_{\epsilon} R_0) }-
\frac{\kappa_{\epsilon}}{\kappa}\frac{\mathbb{k}_{m}(\kappa R_{0})}{\mathbb{i}_m(\kappa_{\epsilon} R_0) }\right]}
{
 \left[\epsilon\Phi_{m}^{(\epsilon)} -\frac{\kappa_{\epsilon}}{\kappa} \frac{\mathbb{k}_m(\kappa R_0)}{\mathbb{i}_m(\kappa_{\epsilon} R_0) }\right]
\left[\Phi_{m}^{(1)}-\frac{\kappa_{\epsilon}}{\kappa}\frac{\mathbb{k}_m(\kappa R_0)}{\mathbb{i}_m(\kappa_{\epsilon} R_0)} \right]}
}{
\frac{1}{\Pi_{m}^{22}}
+   
\frac{\left(\frac{q \frac{\xi}{c}}{\kappa^{2}\Delta^{2}}
\frac{\frac{m}{\kappa_{\epsilon} R_0}}{\mathbb{i}_m(\kappa_{\epsilon} R_0)}
 \right)^2}
 { \left[\epsilon\Phi_{m}^{(\epsilon)} -\frac{\kappa_{\epsilon}}{\kappa} \frac{\mathbb{k}_m(\kappa R_0)}{\mathbb{i}_m(\kappa_{\epsilon} R_0) }\right]
\left[\Phi_{m}^{(1)}-\frac{\kappa_{\epsilon}}{\kappa}\frac{\mathbb{k}_m(\kappa R_0)}{\mathbb{i}_m(\kappa_{\epsilon} R_0)} \right]}}
~.
\end{align}
\label{eq:scattering-coefficients-hollow-cylinder-app}
\end{subequations}
These quantities are defined in terms of expressions which allow us to distinguish the behavior of a shell from that of a full cylinder. Explicitly, we have that
\begin{subequations}
\label{auxiliary-functions}
\begin{equation}
\Phi_{m}^{(x)}=\frac{ x \mu_{m} -\beta_{m}\frac{\kappa_{\epsilon}}{\kappa} \frac{\mathbb{i}_m(\kappa R_i)}{\mathbb{k}_m(\kappa_{\epsilon} R_i)}}
{x \tilde{\beta}_{m}-  \frac{\kappa_{\epsilon}}{\kappa}  \frac{\mathbb{i}_m(\kappa R_i)}{\mathbb{k}_m(\kappa_{\epsilon} R_i)} },
\end{equation}
where we defined the functions
\begin{gather}
\beta_{m}=\frac{1-\frac{\mathbb{k}_m(\kappa_{\epsilon} R_0)}{\mathbb{i}_m(\kappa_{\epsilon} R_0)}\frac{\frac{I_m(\kappa_{\epsilon} R_i)}{I_m(\kappa_{\epsilon} R_0)}}{\frac{K_m(\kappa_{\epsilon} R_i)}{K_m(\kappa_{\epsilon} R_0)}}}
{1- \frac{\frac{I_m(\kappa_{\epsilon} R_i)}{I_m(\kappa_{\epsilon} R_0)}}{\frac{K_m(\kappa_{\epsilon} R_i)}{K_m(\kappa_{\epsilon} R_0)}}},
\;
\tilde{\beta}_{m}
=\frac{1 -\frac{\mathbb{i}_m(\kappa_{\epsilon} R_i)}{\mathbb{k}_m(\kappa_{\epsilon} R_i)}\frac{\frac{I_m(\kappa_{\epsilon} R_i)}{I_m(\kappa_{\epsilon} R_0)}}{\frac{K_m(\kappa_{\epsilon} R_i)}{K_m(\kappa_{\epsilon} R_0)}}}
{1 -\frac{\frac{I_m(\kappa_{\epsilon} R_i)}{I_m(\kappa_{\epsilon} R_0)}}{\frac{K_m(\kappa_{\epsilon} R_i)}{K_m(\kappa_{\epsilon} R_0)}}},
\\
\mu_{m}
= \frac{1- \frac{\mathbb{i}_m(\kappa_{\epsilon} R_i)}{\mathbb{k}_m(\kappa_{\epsilon} R_i)}
\frac{\mathbb{k}_m(\kappa_{\epsilon} R_0)}{ \mathbb{i}_m(\kappa_{\epsilon} R_0) }
\frac{\frac{I_m(\kappa_{\epsilon} R_i)}{I_m(\kappa_{\epsilon} R_0)}}{\frac{K_m(\kappa_{\epsilon} R_i)}{K_m(\kappa_{\epsilon} R_0)}}}
{1- \frac{\frac{I_m(\kappa_{\epsilon} R_i)}{I_m(\kappa_{\epsilon} R_0)}}{\frac{K_m(\kappa_{\epsilon} R_i)}{K_m(\kappa_{\epsilon} R_0)}}}~.
\end{gather}
\end{subequations}
In addition, we introduce

\begin{subequations}
\begin{equation}
\Pi_{m}^{0}=\frac
{1+\frac{\left(\frac{q \frac{\xi}{c}}{\kappa^{2}\Delta^{2}} \frac{\frac{m}{\kappa_{\epsilon}R_i}}{\mathbb{k}_m(\kappa_{\epsilon} R_i)}   \right)^{2}
- 2 \frac{R_0\mathbb{i}_m(\kappa_{\epsilon} R_0)}{R_i\mathbb{k}_m(\kappa_{\epsilon} R_i)}
\epsilon \left(\mu_{m}-\tilde{\beta}_{m}\beta_{m}\right)}{\left[\tilde{\beta}_{m}-\frac{\kappa_{\epsilon}}{\kappa}\frac{\mathbb{i}_m(\kappa R_i)}{\mathbb{k}_m(\kappa_{\epsilon} R_i)}\right] \left[\epsilon\tilde{\beta}_{m} -\frac{\kappa_{\epsilon}}{\kappa}\frac{\mathbb{i}_m(\kappa R_i)}{\mathbb{k}_m(\kappa_{\epsilon} R_i)}\right]}}
{1+\frac{\left(\frac{q \frac{\xi}{c}}{\kappa^{2}\Delta^{2}}\frac{\frac{m}{\kappa_{\epsilon} R_{i}}}{\mathbb{k}_m(\kappa_{\epsilon} R_i)}\right)^2
-\frac{R_{0}}{R_{i}} \frac{\mathbb{i}_m(\kappa_{\epsilon} R_0)}{\mathbb{k}_m(\kappa_{\epsilon} R_i)} \epsilon(\mu_{m}-\tilde{\beta}_{m}\beta_{m})}
{\left[\tilde{\beta}_{m} - \frac{\kappa_{\epsilon}}{\kappa}\frac{\mathbb{i}_{m}(\kappa R_{i})}{\mathbb{k}_m(\kappa_{\epsilon} R_i)}\right]
\left[\epsilon\tilde{\beta}_{m} -\frac{\kappa_{\epsilon}}{\kappa }\frac{\mathbb{i}(\kappa R_{i})}{\mathbb{k}_m(\kappa_{\epsilon} R_i)} \right]}}
\end{equation}
and also 

\begin{widetext}
\begin{equation}
\Pi_{m}^{ij}=\frac
{1+\frac{\left(\frac{q \frac{\xi}{c}}{\kappa^{2}\Delta^{2}} \frac{\frac{m}{\kappa_{\epsilon}R_i}}{\mathbb{k}_m(\kappa_{\epsilon} R_i)}   \right)^{2}-2 \frac{R_0\mathbb{i}_m(\kappa_{\epsilon} R_0)}{R_i\mathbb{k}_m(\kappa_{\epsilon} R_i)}\epsilon \left(\mu_{m}-\tilde{\beta}_{m}\beta_{m}\right)}{\left[\tilde{\beta}_{m}-\frac{\kappa_{\epsilon}}{\kappa}\frac{\mathbb{i}_m(\kappa R_i)}{\mathbb{k}_m(\kappa_{\epsilon} R_i)}\right] \left[\epsilon\tilde{\beta}_{m} -\frac{\kappa_{\epsilon}}{\kappa}\frac{\mathbb{i}_m(\kappa R_i)}{\mathbb{k}_m(\kappa_{\epsilon} R_i)}\right]}}
{1 +
 \frac{\left(\frac{q \frac{\xi}{c}}{\kappa^{2}\Delta^{2}} \frac{\frac{m}{\kappa_{\epsilon}R_i}}{\mathbb{k}_m(\kappa_{\epsilon} R_i)}   \right)^{2}}{\left[\tilde{\beta}_{m}-\frac{\kappa_{\epsilon}}{\kappa}\frac{\mathbb{i}_m(\kappa R_i)}{\mathbb{k}_m(\kappa_{\epsilon} R_i)}\right] \left[\epsilon\tilde{\beta}_{m} -\frac{\kappa_{\epsilon}}{\kappa}\frac{\mathbb{i}_m(\kappa R_i)}{\mathbb{k}_m(\kappa_{\epsilon} R_i)}\right]}
 \frac{
 \left[\beta_{m}-\frac{\kappa_{\epsilon}}{\kappa}\frac{\mathbb{x}^{(i)}_m(\kappa R_0)}{\mathbb{i}_m(\kappa_{\epsilon} R_0)} \right]
 \left[\epsilon\beta_{m}-\frac{\kappa_{\epsilon}}{\kappa}\frac{\mathbb{x}^{(j)}_m(\kappa R_0)}{\mathbb{i}_m(\kappa_{\epsilon} R_0)}\right]
}{
\left[\Phi_{m}^{(1)}-\frac{\kappa_{\epsilon}}{\kappa}\frac{\mathbb{x}^{(i)}_m(\kappa R_0)}{\mathbb{i}_m(\kappa_{\epsilon} R_0)}\right] 
\left[\epsilon \Phi_{m}^{(\epsilon)} -\frac{\kappa_{\epsilon}}{\kappa} \frac{\mathbb{x}^{(j)}_m(\kappa R_0)}{\mathbb{i}_m(\kappa_{\epsilon} R_0)} \right]
}}
~.
\end{equation}
\end{widetext}
\label{auxiliary-functions-2}
\end{subequations}
In the last expression, the superscript identifies $\mathbb{x}^{(1)}_{m}\equiv\mathbb{i}_{m}$ and $\mathbb{x}^{(2)}_{m}\equiv\mathbb{k}_{m}$.
As $R_{i}\to 0$, the functions $\Phi_{m}^{(x)}$, $\tilde{\beta}_{m}$, $\beta_{m}$, $\mu_{m}$, and $\Pi_{m}^{ij}$ tend to one. Consequently, the expressions of the corresponding scattering coefficients for the full cylinder can be directly derived from Eqs.~\eqref{eq:scattering-coefficients-hollow-cylinder-app}.

\section{Additional information on the asymptotic analysis}
\label{app:asymptotics}

For completeness, we provide additional information about the derivations of the expressions presented in Sec.~\ref{sec:zero-temperature-interaction}. Since the slab-limit has been discussed several times in the literature~\cite{Intravaia2011,Buhmann2012,Buhmann2012a}, our main focus here is on the thin-wire limit of the Casimir-Polder interaction, i.e. the asymptotic expression resulting from a simplification of \cref{eq:cp-interaction-dimensionless-text} for $s\ll 1$. This analysis is more conveniently done using the new variables given in Eqs.~\eqref{change-of-variables}, since in this case we can identify that the dominant contribution to the free energy arises for $\eta \approx 1$ (see main text) and $\zeta \in (0,1)$. 

\subsection{Approximate form of the scattering coefficients}

Defining $\chi=R_{i}/R_{0}$ and using 
that $\eta \approx 1$, and $\zeta \in (0,1)$, the auxiliary functions defined in Eqs.~\eqref{auxiliary-functions} simplify for $s\ll 1$ as follows
\begin{subequations}
\begin{align}
\Phi_{m}^{(\epsilon)}
&\sim
\begin{cases}
\frac{1-\chi^{2}+\Delta^{2}}
{1+\Delta^{2}} 
& m= 0
\\
\\
\frac{ \left(1+\frac{1}{\Delta^{2}}\right) +\frac{1 +\chi^{2m}}
{1 -\chi^{2m}}\left(1+\frac{\zeta^{2}}{\Delta^{2}}\right)}
{\left(1+\frac{1}{\Delta^{2}}\right)\frac{1 +\chi^{2m}}
{1 -\chi^{2m}}+\left(1+\frac{\zeta^{2}}{\Delta^{2}}\right)}
& m\neq 0~,
\end{cases}
\end{align}

\begin{align}
\Phi_{m}^{(1)}
&\sim
\begin{cases}
1& m= 0
\\
\\
\frac{ 1 +\frac{1 +\chi^{2m}}
{1 -\chi^{2m}}\left(1+\frac{\zeta^{2}}{\Delta^{2}}\right)}
{\frac{1 +\chi^{2m}}
{1 -\chi^{2m}}+\left(1+\frac{\zeta^{2}}{\Delta^{2}}\right)}
& m\neq 0~.
\end{cases}
\end{align}
The other quantities are relevant only for $m\neq 0$. We obtain $\mu_{m}\sim 1$ and
\begin{equation}
\beta_{m}\sim\tilde{\beta}_{m}\sim \frac{1 +\chi^{2m}}{1 -\chi^{2m}}\equiv b_{m}~.
\end{equation}
Moreover, in the thin-wire limit, we find that
\begin{equation}
\Pi_{m}^{0}\sim\frac
{1+\frac{\frac{\zeta^{2} (1-\zeta^{2})}{\Delta^{4}}+2\left(1+\frac{1}{\Delta^{2}}\right) \left(1-b_{m}^{2}\right)}
{\left[b_{m}+\left(1+\frac{\zeta^{2}}{\Delta^{2}}\right)\right] 
\left[\left(1+\frac{1}{\Delta^{2}}\right) b_{m}+\left(1+\frac{\zeta^{2}}{\Delta^{2}}\right)\right]}}
{1+\frac{\frac{\zeta^{2} (1-\zeta^{2})}{\Delta^{4}}+\left(1+\frac{1}{\Delta^{2}}\right) \left(1-b_{m}^{2}\right)}
{\left[b_{m}+\left(1+\frac{\zeta^{2}}{\Delta^{2}}\right)\right] 
\left[\left(1+\frac{1}{\Delta^{2}}\right) b_{m}+\left(1+\frac{\zeta^{2}}{\Delta^{2}}\right)\right]}}
\end{equation}
and also

\begin{widetext}
\begin{equation}
\Pi_{m}^{ij}\sim\frac
{1+\frac{\frac{\zeta^{2} (1-\zeta^{2})}{\Delta^{4}}+2\left(1+\frac{1}{\Delta^{2}}\right) \left(1-b_{m}^{2}\right)}
{\left[b_{m}+\left(1+\frac{\zeta^{2}}{\Delta^{2}}\right)\right] 
\left[\left(1+\frac{1}{\Delta^{2}}\right) b_{m}+\left(1+\frac{\zeta^{2}}{\Delta^{2}}\right)\right]}}
{1 + \frac{\frac{\zeta^{2} (1-\zeta^{2})}{\Delta^{4}}}
{\left[b_{m}+\left(1+\frac{\zeta^{2}}{\Delta^{2}}\right)\right] \left[\left(1+\frac{1}{\Delta^{2}}\right) b_{m}+\left(1+\frac{\zeta^{2}}{\Delta^{2}}\right)\right]}
 \frac{
 \left[b_{m}+(-1)^{i}\left(1+\frac{\zeta^{2}}{\Delta^{2}}\right)\right]
 \left[\left(1+\frac{1}{\Delta^{2}}\right) b_{m}+(-1)^{j}\left(1+\frac{\zeta^{2}}{\Delta^{2}}\right)\right]
}{
\left[\Phi_{m}^{(1)}+(-1)^{i}\left(1+\frac{\zeta^{2}}{\Delta^{2}}\right)\right] 
\left[\left(1+\frac{1}{\Delta^{2}}\right)  \Phi_{m}^{(\epsilon)}+ (-1)^{j}\left(1+\frac{\zeta^{2}}{\Delta^{2}}\right)\right]
}}
\quad (m\neq 0)~.
\end{equation}
\end{widetext}
\end{subequations}
Importantly we see that, in the thin-wire limit for $m\neq0$, all the auxiliary functions depend on the variable $\eta$, a property that is inherited by the scattering coefficients in \cref{eq:scattering-coeff-thin-wire-mnot0}.
When inserted in Eqs.~\eqref{eq:scattering-coefficients-hollow-cylinder-app}, the previous approximations indeed lead to the expressions in Eqs.~\eqref{eq:r0NN} and \eqref{eq:scattering-coeff-thin-wire-mnot0}.

\subsection{The Casimir-Polder free energy in the thin-wire limit}
\label{app:thin-wire}

As discussed before \cref{eq:relative-penetration-depth-materials-zero-frequency}, depending on the material, the behavior of $\Delta$ can be quite different, deeply affecting the asymptotic behavior of the Casimir-Polder free energy. This is particularly evident in the expression for $r_{0}^{\mathrm{NN}}$. Indeed, for dielectrics the function $\Delta$ has a positive lower bound for $\xi\to 0$ and in the limit $s\ll 1$, compatibly with the condition in \cref{eq:condition-dielectrics}, this allows to neglect the logarithmic term in \cref{eq:r0NN}, obtaining

\begin{align}
\Delta > \Delta_{0}>0 \quad \Rightarrow r_{0}^{\mathrm{NN}}
\sim -\frac{(\eta s)^{2} \ln(\eta s \tilde{\gamma}_{\mathrm{E}})}{2\Delta^{2}}
~.
\end{align}
The previous simplification is not possible for conductors, since $\Delta$ can go to zero. Specifically, in the limit where the expressions in \cref{eq:relative-penetration-depth-materials-zero-frequency} are valid, we obtain for non-dissipative conductors
\begin{equation}
\Delta\sim  \zeta\frac{\lambda_{p}}{R_{0}}\eta s 
\quad\Rightarrow 
r_{0}^{\mathrm{NN}}\sim \frac{(1-\chi^{2})\frac{R^{2}_{0}}{2 \lambda_{p}^2}\ln(\eta s \tilde{\gamma}_{\mathrm{E}} )}
{(1-\chi^{2})\frac{R^{2}_{0}}{2 \lambdabar_{p}^2}\ln(\eta s \tilde{\gamma}_{\mathrm{E}})- \zeta^{2}}~,
\end{equation}
while for ohmic conductors we obtain
\begin{equation}
\Delta\sim  \sqrt{\zeta\frac{\lambda_{D}}{R_{0}}\eta s} 
\quad\Rightarrow 
r_{0}^{\mathrm{NN}}\sim \frac{(1-\chi^{2})\frac{R_{0}}{2 \lambdabar_{D}}(\eta s)\ln(\eta s \tilde{\gamma}_{\mathrm{E}})}
{(1-\chi^{2})\frac{R_{0}}{2 \lambdabar_{D}}(\eta s)\ln(\eta s \tilde{\gamma}_{\mathrm{E}})-\zeta}~.
\end{equation}

An analysis of \cref{eq:cp-interaction-dimensionless-text} allows to conclude that, for $s\ll 1$ the leading contribution to the Casimir-Polder free energy can be written as
\begin{widetext}
\begin{align}
\label{eq:cp-interaction-zero-temperature-simplified}
\mathcal{F}
\sim
&
 - 
\frac{\hbar c}{4 \pi^3}  \frac{1}{L^4}
\int\limits_{0}^{\infty} \dd{\eta}
\int\limits_0^{1} \dd{\zeta} \; \frac{\alpha\left(\zeta\frac{c}{R_{0}}\eta s\right)}{\epsilon_{0}}\frac{r_{0}^{\mathrm{NN}}}{-\ln(\eta s \gamma^{\prime}_{\mathrm{E}} )}\frac{1 + (1-\zeta^{2}) \mathbb{k}^{2}_0(\eta)}{\sqrt{1-\zeta^2}}
  \eta^{3} K_0^2(\eta )
\nonumber\\
&
- 
\frac{\hbar c}{2 \pi^3}  \frac{1}{L^4}
\int\limits_{0}^{\infty} \dd{\eta}
\int\limits_0^{1} \dd{\zeta} \; \frac{\alpha\left(\zeta\frac{c}{R_{0}}\eta s\right)}{\epsilon_{0}}
\\
&\hspace{2cm}\times
   \Bigg\{ \frac{ 1 + (1-\zeta^{2})  \left[   \frac{1}{\eta^{2}} +   \mathbb{k}^{2}_1(\eta) \right]}{\sqrt{1-\zeta^2}}
   r_{1}^{\mathrm{NN}}  
 -  \frac{\zeta^{2} \left[   \frac{1}{\eta^{2}} +   \mathbb{k}^{2}_1(\eta) \right]}{\sqrt{1-\zeta^2}} 
 r_{1}^{\mathrm{MM}} 
+ 4\zeta \frac{ 1 }{\eta} \mathbb{k}_1 (\eta ) 
r_{1}^{\mathrm{MN}}
\Bigg\} \eta^{3}2\left(\frac{\eta s}{2}\right)^{2} K_1^{2}(\eta )~.
\nonumber
\end{align}
\end{widetext}
We notice that the two integrals above correspond to the terms for $m=0,1$ of \cref{eq:cp-interaction-dimensionless-text}. Higher orders can be neglected due to their scaling as $s^{2m}$ for $m\ge 2$. 
The same orders also provide the dominant contribution in the thermal regime. According to the discussion at the beginning of \cref{sec:finite-temperature-interaction}, setting $\zeta=0$ in the integrands of \cref{eq:cp-interaction-zero-temperature-simplified} and dividing it by  $\eta c/L$, we can readily write the $n=0$ (zeroth Matsubara) term for the nonzero temperature Casimir-Polder free energy [see \cref{eq:cp-interaction-dimensionless-finite-temperature}] 
\begin{align}
\label{eq:cp-interaction-finite-temperature-simplified}
\mathcal{F}
\sim
&
 - 
\frac{k_{B}T}{4 \pi^2 \epsilon_{0}}  \frac{\alpha_{0}}{L^3}
\int\limits_{0}^{\infty} \dd{\eta}
\frac{r_{0}^{\mathrm{NN}}}{-\ln(\eta s \gamma^{\prime}_{\mathrm{E}} )}\left[1 + \mathbb{k}^{2}_0(\eta)\right]  \eta^{2} K_0^2(\eta )
\nonumber\\
&
- 
\frac{k_{B}T}{4\pi^2 \epsilon_{0}}  \frac{\alpha_{0}}{L^3}s^{2}
\int\limits_{0}^{\infty} \dd{\eta}
  \left[1 +    \frac{1}{\eta^{2}} +   \mathbb{k}^{2}_1(\eta) \right]r_{1}^{\mathrm{NN}}  \eta^{4} K_1^{2}(\eta )~.
\end{align}
For dielectrics, due to the behavior of $r_{0}^{\mathrm{NN}}$ discussed above, both integrals in \cref{eq:cp-interaction-zero-temperature-simplified,eq:cp-interaction-finite-temperature-simplified} scale with $s^{2}$. This scaling is responsible for the different power laws discussed in the main text (see ~\cref{sec:dielectric-hollow-core-cylinder}). 

A different picture emerges for conductors: The $m=0$ term (the first line in \cref{eq:cp-interaction-zero-temperature-simplified,eq:cp-interaction-finite-temperature-simplified}) provides the dominant contribution. 
Notice, in particular, that the respective integrands considered above contain some functions depending only on $\eta s$ and others depending only on $\eta$. In this specific case, due to the fact that $s\ll 1$, the former vary slower with $\eta$ than the latter, which also features a positive peak for $\eta \approx 1$. This suggests the following strategy for an approximate evaluation of the integrals in $\eta$ in the first line in \cref{eq:cp-interaction-zero-temperature-simplified,eq:cp-interaction-finite-temperature-simplified}. The integrand has the general form
\begin{align}
\int\limits_{0}^{\infty} \dd{\eta} \; g(s\eta)f(\eta)
&\stackrel{y=\eta s}{=}
\int\limits_{0}^{\infty} \dd{y}  \; g(y) \frac{e^{\ln\left[f\left(\frac{y}{s}\right)\right]}}{s}~,
\end{align}
where the function $g$ is smooth, while the function $f$ exhibits a peak at $\eta=\eta_{0}$.
We then perform a saddle-point approximation, which consists in expanding the exponent in the above expression around $y_{0}=\eta_{0}s$, using that $f'\left(\frac{y_{0}}{s}\right)=0$. We have 

\begin{equation}
\label{saddle-point}
\frac{1}{s}e^{\ln\left[f\left(\frac{y}{s}\right)\right]}\sim e^{\ln[f(\eta_{0})]} \frac{e^{\frac{f''(\eta_{0}) (y-\eta_{0} s)^2}{2 s^2 f(\eta_{0})}}}{s}~.
\end{equation}
Given that $f''\left(\frac{y_{0}}{s}\right)<0$, the expression at \cref{saddle-point} is proportional to a Gauss function with maximum in $y=\eta_{0} s$ and width $s \sqrt{f(\eta_{0})}>0$. 
In the limit $s\ll 1$ the Gauss function becomes proportional to a Dirac delta function so that for the original integral we can write 
\begin{align}
\int\limits_{0}^{\infty} \dd{\eta}  \; g(s\eta)f(\eta)\stackrel{s\ll 1}{\sim}
g(s\eta_{0}) \int\limits_{0}^{\infty} \dd{\eta} \; f(\eta)~.
\end{align}
This approach can be used for determining the asymptotic behavior of ${\cal F}$ both in the regime $R_0 \ll L \ll \lambda_a$ and in the limit $L \gg \lambda_a$ (for instance, see \cref{eq:conductor-nonretarded-start,eq:conductor-retarded-start}). 

In order to illustrate the role of this approach in the analysis of the main text, it is interesting to consider a few examples. 
Let us consider \cref{eq:conductor-retarded-start} in the superconducting case. If we replace the function $\Delta$ with its behavior at low frequency given in \cref{eq:relative-penetration-depth-materials-zero-frequency}, the expression describing the  free energy can be calculated using the following steps. We first carry out the saddle-point approximation 
\begin{align}
\mathcal{F}
&\sim
- 
\frac{\hbar c}{8 \pi^3 } \frac{\frac{\alpha_{0}}{\epsilon_{0}}}{L^4}
\int\limits_{0}^{\infty} \dd{\eta}
\int\limits_0^{1} \dd{\zeta} \; 
 \frac{\frac{1 + (1-\zeta^{2}) \mathbb{k}^{2}_0(\eta)}{\sqrt{1-\zeta^2}}\eta^{3}K_0^2(\eta ) }
{\frac{\left(\zeta\frac{\lambdabar_{p}}{R_{0}}\right)^{2}}{(1-\chi^{2})}-\frac{\ln(\eta s \tilde{\gamma}_{\mathrm{E}})}{2} }
\nonumber\\
&\sim
- 
\frac{\hbar c}{8 \pi^3} \frac{\frac{\alpha_{0}}{\epsilon_{0}}}{L^4}
\int\limits_0^{1} \dd{\zeta} \; 
 \frac{\int\limits_{0}^{\infty} \dd{\eta}\frac{1 + (1-\zeta^{2}) \mathbb{k}^{2}_0(\eta)}{\sqrt{1-\zeta^2}}\eta^{3}K_0^2(\eta ) }
{\frac{\left(\zeta\frac{\lambdabar_{p}}{R_{0}}\right)^{2}}{(1-\chi^{2})}-\frac{\ln(\eta_{[3]} s \tilde{\gamma}_{\mathrm{E}})}{2} }
~.
\end{align}
The resulting integrand in $\eta$ can be evaluated analytically and we have
\begin{align}
\label{superconductor-retarded-app}
\mathcal{F}
&\sim
- 
\frac{\hbar c}{8 \pi^3} \frac{\frac{\alpha_{0}}{ \epsilon_{0}}}{L^4}
\int\limits_0^{1} \dd{\zeta} \; 
 \frac{ \frac{1-\frac{2}{3}\zeta^{2}}{\sqrt{1-\zeta ^2}}}
{\frac{\left(\zeta\frac{\lambdabar_{p}}{R_{0}}\right)^{2}}{(1-\chi^{2})}-\frac{\ln(\eta_{[3]} s \tilde{\gamma}_{\mathrm{E}})}{2} }
~.
\end{align}
The integration over $\zeta$ then leads to the expression in \cref{eq:superconductor-retarded-text} of the main text.

Another example is the case of the ohmic conductor. Starting from \cref{eq:conductor-retarded-start} in the main text,
we obtain
\begin{align}
\label{eq:conductor-retarded-start-app}
\mathcal{F}
&\sim
- 
\frac{\hbar c}{8 \pi^3 \epsilon_{0}} \frac{\alpha_{0}}{L^4} s
\int\limits_{0}^{\infty} \dd{\eta}
\int\limits_0^{1} \dd{\zeta} \; 
 \frac{\frac{1 + (1-\zeta^{2}) \mathbb{k}^{2}_0(\eta)}{\sqrt{1-\zeta^2}}\eta^{4}K_0^2(\eta ) }
{\frac{\zeta\frac{\lambdabar_{D}}{R_{0}}}{(1-\chi^{2})}-\frac{(\eta s)\ln(\eta s \tilde{\gamma}_{\mathrm{E}})}{2} }
\nonumber\\
&\sim
- 
\frac{\hbar c}{8 \pi^3 \epsilon_{0}} \frac{\alpha_{0}}{L^4} s
\int\limits_0^{1} \dd{\zeta} \; 
 \frac{\int\limits_{0}^{\infty} \dd{\eta}\frac{1 + (1-\zeta^{2}) \mathbb{k}^{2}_0(\eta)}{\sqrt{1-\zeta^2}}\eta^{4}K_0^2(\eta ) }
{\frac{\zeta\frac{\lambdabar_{D}}{R_{0}}}{(1-\chi^{2})}-\frac{(\eta_{[4]} s)\ln\left(\eta_{[4]} s \tilde{\gamma}_{\mathrm{E}}\right)}{2} }
\nonumber\\
&=
- 
\frac{\hbar c}{8 \pi^3 \epsilon_{0}}\frac{\alpha_{0}R_{0}^{2}}{L^5\lambdabar_{D}}(1-\chi^{2})
\int\limits_0^{1} \dd{\zeta} \; 
 \frac{-\frac{9 \pi ^2 \left(5 \zeta ^2-8\right)}{512 \sqrt{1-\zeta ^2}}}
{\zeta-\frac{(\eta_{[4]} s)\ln\left(\eta_{[4]} s \tilde{\gamma}_{\mathrm{E}}\right)}{2
\frac{\lambdabar_{D}}{R_{0}(1-\chi^{2})}} }
~.
\end{align}
The last integral can also be performed analytically but its expression is not very transparent and will not be reported here.
An approximate expression, under the condition of \cref{condition-conductor-retarded-full}, is reported in \cref{eq:conductor-retarded}.

Finally, using the expression for $r_{0}^{\mathrm{NN}}$ given in \cref{eq:r0NN}, within the nonretarded region of the thin-wire limit for a conductor, the first line of \cref{eq:cp-interaction-zero-temperature-simplified} can be written as
\begin{align}
\label{eq:conductor-nonretarded-appendix}
\mathcal{F}&\sim
- 
\frac{\hbar}{8 \pi^3} \frac{R_{0}s}{L^4}
\int\limits_0^{\infty} \dd{\eta}
\int\limits_0^{\infty} \dd{\xi} 
\frac{\frac{\alpha\left(\ii \xi\right)}{\epsilon_0}\eta^{4} [1 + \mathbb{k}^{2}_0(\eta)]K_0^2(\eta )}
{\frac{\Delta^{2}\left(\ii \xi \right)}{1-\chi^{2}}-\frac{(\eta s)^{2}}{2} \ln(\eta s \gamma^{\prime}_{\rm E})}~.
\end{align}
Using the saddle-point approximation we obtain in this case \cref{eq:conductor-nonretarded-start}. It is worth to highlight the role played by the polarizability in the convergence of the $\xi$-integral when the ohmic expression for $\Delta$ is employed. The expression in \cref{eq:conductor-nonretarded-appendix} is obtained by performing a partial integration in $\xi$ and by realizing that $\partial_{\xi}\alpha\left(\ii \xi\right)$ takes on appreciable values only for $\xi \sim \omega_{a}$. In the same spirit of the saddle point approximation, we can therefore replace $\xi \sim \omega_{a}$ in the remaining part of the integrand. The remaining integration, only involving $\partial_{\xi}\alpha\left(\ii \xi\right)$, gives $\alpha_{0}$.

\section{Details on the numerical evaluation}
\label{app:numerical-evaluation-details}

In this appendix, we provide additional details on the numerical evaluation of the Casimir-Polder interaction in systems involving a cylindrical structure discussed in \cref{sec:numerical-evaluation}. 
In addition, we also introduce the necessary modifications to treat finite-temperature effects. 

As discussed in the main text, we treat the combined integration and summation in \cref{eq:cp-interaction-dimensionless-text,eq:cp-interaction-dimensionless-finite-temperature} with a multivariate Gaussian quadrature scheme. 
Specifically, we use a tensor product quadrature on a three dimensional grid, where the weights and nodes in each dimension correspond to a $1\mathrm{D}$ Gaussian quadrature rule adapted to the particular behavior of the integrand along a given direction. 
This choice depends on several factors. 
In the case of finite intervals, the standard rules based on the inner products of classical orthogonal polynomials are typically over the range $[-1,1]$. 
Ideally, known integrable singularities should be incorporated into the quadrature rule to improve accuracy. For example, embedding their behavior directly into the weight function allows for the efficient determination of optimally adapted nodes and weights (a detailed description specific to our system is provided in the section below).
In addition, when dealing with infinite or semi-infinite intervals, the asymptotic decay behavior of the integrand needs to be taken into account.   
All computations are performed with the Julia programming language \cite{Bezanson2017}.

\subsection{Zero temperature interaction}

\subsubsection{Integral over $\zeta$}
As noted in the main text, the integration over $\zeta$ in \cref{eq:cp-interaction-dimensionless-text} contains an integrable square root singularity at $\zeta=1$. 
In principle, this could be treated directly by a specialized half-range Chebyshev-Gauss quadrature rule based on Chebyshev-like polynomials defined on the range $[0,1]$~\cite{Huybrechs2010}. 
Here, we instead transform the integration interval from $[0,1]$ to the standard interval $[-1,1]$ via the change of variable $\tilde{\zeta} = 2\zeta-1$. 
Specifically, we find that
\begin{equation}
\frac{1}{\sqrt{1-\zeta^{2}}}
=
\frac{2}{\sqrt{1-\tilde{\zeta}} \sqrt{3+\tilde{\zeta}}}
\end{equation}
has singularities at $\tilde{\zeta}=-3$ and $\tilde{\zeta}=1$, 
where only the latter is of relevance to the numerical evaluation of the integral.  
This can be treated via an $N_{\zeta}$ point Gauss-Jacobi quadrature for the weight function $(1-x)^{\alpha}(1+x)^{\beta}$ with $\alpha=-1/2$ and $\beta=0$, which is readily available in most numerical software libraries. 

\subsubsection{Integral over $\eta$}
\label{app:sec:numerics-eta-integration}

In order to efficiently compute the $\eta$ integral over the semi-infinite interval $[0,\infty)$, it is useful to analyze the asymptotic decay behavior of the integrand for large $\eta$. 
Apart from algebraic factors, the integrand nontrivially depends on $\eta$ via the polarizability, the logarithmic derivatives of the Bessel functions, which also enter into the scattering coefficients, and the Kernel function $\mathcal{K}_{m}(\eta,s)$. 
The polarizability always vanishes as $\eta^{-2}$ for large values of $\eta$, while the logarithmic derivatives $\mathbb{i}_{m}$ and $-\mathbb{k}_{m}$ approach unity for large arguments (see \cref{eq:modified-bessel-logarithmic-derivatives-asymptotics}).
The dominant influence results from the kernel function $\mathcal{K}_{m}$ in \cref{eq:cp-kernel-function}
and is determined by the exponential large-argument scaling of the modified Bessel functions in \cref{eq:modified-bessel-argument-asymptotics}.
In particular, in this expression both exponentially growing and decreasing modified Bessel functions combine in such a way as to lead to an overall decay, making the integral convergent. 

As already discussed in the main text, it is advantageous for the numerical evaluation of \cref{eq:cp-interaction-dimensionless-text} to make the exponentially decreasing nature of the integrand explicit. 
To this end, we utilize the exponentially scaled modified Bessel functions $\tilde{I}_m(x)=I_m(x)e^{-x}$ and $\tilde{K}_m(x)=K_m(x)e^{x}$.  
Implementations of these functions are available from standard numerical libraries~\cite{Amos1986}. 
From the identities~\cite{Abramowitz71}
\begin{equation}
\begin{aligned}
2 I'_{m}(x) &= I_{m+1}(x) + I_{m-1}(x)
\\
2 K'_{m}(x) &= -\left[K_{m+1}(x) + K_{m-1}(x)\right]
\end{aligned}
\end{equation} 
we can similarly define the exponentially scaled derivatives $\tilde{I}'_m(x)=I'_m(x)e^{-x}$ and $\tilde{K}'_m(x)=K'_m(x)e^{x}$ which exhibit the same scaling as the original functions. 
In our implementation, the values of the logarithmic derivatives $\mathbb{i}_{m}$ and $\mathbb{k}_{m}$ thus remain unaffected when replacing the modified Bessel functions by their exponentially scaled counterparts.  
Meanwhile, performing these replacements in \cref{eq:cp-kernel-function} leads to the exponentially scaled kernel function $\tilde{\mathcal{K}}_{m}$. 
The latter is related to the original function by an explicit multiplicative factor $\ee^{-2\eta}$, as pointed out in \cref{eq:kernel-function-exponentially-scaled}. 
With the further change of variable $\tilde{\eta} = 2 \eta$, the integral over $\eta$ can thus be efficiently evaluated by an $N_{\eta}$-point Gauss-Laguerre quadrature rule.

\subsubsection{Infinite series over $m$}

In order to treat the summation over $m$ in \cref{eq:cp-interaction-dimensionless-text} on the same footing as the integrations, we employ a generalized notion of Gaussian quadrature.  
While the latter is commonly used for the evaluation of integrals, it can also be applied in the context of infinite series. 
This results in a Gaussian summation scheme effectively corresponding to a quadrature with a discrete measure~\cite{Engblom2006,Monien2010}. 
Similar to the integration on a semi-infinite interval, also in the discrete case it is essential to identify the asymptotic behavior of the summand for large orders responsible for eventually making the series converge. 
This is again determined by the combinations of modified Bessel functions and their logarithmic derivatives. 
From Eqs.~\eqref{eq:bessel-bounds}, we have that $\mathbb{i}_{m}$ and $\mathbb{k}_{m}$ scale at most linearly with $m$ for $m\gg 1$ for positive arguments~\cite{Amos74,Yang17a}, a property that is inherited by the scattering coefficients (see \cref{eq:scattering-coefficients-hollow-cylinder-app,auxiliary-functions,auxiliary-functions-2}).
As discussed in the main text, the kernel function and thus the whole summand decreases exponentially for large orders, with a rate of $\mu=2\log{\left(1+\frac{1}{s}\right)}$ (c.p. \cref{eq:cp-kernel-function-large-m-asymptotics}). 
The previous argument is based on a pointwise asymptotic expansion for large orders.   
Additionally, the bounds reported in \cref{eq:bessel-bounds} allow to show that for $\forall\;\eta,s\ge 0$ and $m \ge 1$ the kernel function fulfills~\cite{Baricz10,Segura21}
\begin{equation}
\mathcal{K}_{m}(\eta,s)<\frac{\eta^{3}e^{-2m \ln\left(1+\frac{1}{s}\right)}}{2\sqrt{\left(m-\frac{1}{2}\right)^{2}+(s \eta)^{2}}}~,
\end{equation}
indicating that the pointwise expansion saturates the previous bound for $m \to \infty$. 
Similarly, starting from a uniform expansion of the Bessel functions for $m\to\infty$~\cite{Abramowitz71}, it can be shown that the pointwise expansion indeed constitutes the limiting factor for the convergence of the sum for $s\gg 1$. 

All the previous results show that in the limit of $s\ll 1$, the sum rapidly decays with $m$, suggesting that a direct summation is close to optimal.  These considerations also allow to identify a suitable generalized quadrature rule for the sum in \cref{eq:cp-interaction-dimensionless-text} in the large-aspect ratio limit ($s\gg 1$). 
Specifically, a Gaussian summation on the basis of modified discrete Laguerre (MDL) polynomials for infinite series of the form
\begin{equation}
\primesum_{n=0}^{\infty} h f(n h)e^{-\mu n h }
\approx 
\sum_{i=1}^{N_{n}}w_{i} f(x_{i})
\label{eq:MDL-gaussian-summation}
\end{equation}
was developed in Ref.~\cite{Xu2021}. 
In the above expression $\mu$ and $h$ are real quantities characterizing the sum on the left of \cref{eq:MDL-gaussian-summation}, $N_{n}$ is the number of quadrature points in the sum on the right, while $w_{i}$ and $x_{i}$ are real numbers indicating, respectively, the weights and nodes of the quadrature rule~\cite{Press2007}.
Based on the three-term recurrence relations given in \cite{Xu2021}, we have implemented the calculation of the nodes and weights necessary for the MDL quadrature via the Golub-Welsch algorithm~\cite{Golub1969}. 

In our case, the index of the first sum coincides with the order $m\ge0$ of the Bessel functions (c.p.\ \cref{eq:cp-interaction-dimensionless-text}). 
If we set $h = 1$, we can thus make use of the MDL quadrature rule for the sum in \cref{eq:cp-interaction-dimensionless-text}. 
We note that while the asymptotic exponential decay with $m$ is again encoded in the Kernel function, it does not appear explicitly. 
This is in contrast to the $\eta$ integration, where a formulation in terms of the scaled versions of the modified Bessel functions allows for the direct use of a Gauss-Laguerre quadrature rule by simply omitting the exponential factor $\ee^{-\tilde{\eta}}$ (c.p.\ \cref{app:sec:numerics-eta-integration}). 
Here, we instead employ an implicit quadrature rule where the exponential decay with respect to $m$ is incorporated into the weights. 
Furthermore, particular attention has to be devoted to the fact that, in the MDL implementation, the order of the Bessel functions becomes a continuous real variable instead of an integer number. 
While implementations of Bessel functions for arbitrary real-valued orders exist, it sometimes occurs that the values required by the quadrature rule are such that they lead to under- or overflow in standard double precision routines. 
Similar problems occur in a direct evaluation of the integrand for very large orders or arguments. 
In those cases, we switch to arbitrary precision arithmetic. 
Since the total number of points that needs to be evaluated is generally small, we have found the additional overhead thus incurred to be negligible in practice. 
To further optimize this point, one could instead switch to an asymptotic expansion of the integrand for extreme values of Bessel function arguments and/or orders.  
In practice, for all the numerical evaluations in this work, we switch to the MDL scheme for the sum over $m$ if $s > 1$, otherwise using direct summation. 
In the latter case, we generally find it more efficient to replace the fixed-point rules for the integrals over $\eta$ and $\zeta$ with an adaptive 2D quadrature routine based on Genz-Malik quadrature \cite{Genz1980,Johnson2017}.

\subsection{Finite temperature interaction}

We shall briefly summarize the necessary modifications to include finite-temperature effects in the numerical evaluation according to \cref{eq:cp-interaction-finite-temperature,eq:cp-interaction-dimensionless-finite-temperature}.
With respect to the zero temperature case, the major difference in the mathematical expressions arises from the formal replacement  of the integral over imaginary frequencies $\xi$ to an infinite sum over the Matsubara frequencies $\xi_{n}$
(see \cref{eq:zero-to-finite-temperature-replacement}). 
This case essentially corresponds to one of the applications of the MDL summation technique originally proposed in~\cite{Xu2021}, namely the evaluation of finite-temperature Casimir forces. 
From our previous discussion, we thus expect an exponential decay of the sum over $n$ with a rate proportional to the temperature. 
To make this behavior explicit, we introduce the following change of variables starting from \cref{eq:variable-change-finite-temperature}  
\begin{equation}
\eta_{T} = \eta - n \frac{L}{\lambdabar_{T}}~,
\end{equation}
thereby shifting the limit of the $\eta_{T}$ integration to the interval $[0,\infty)$. 
Consequently, from \cref{eq:kernel-function-exponentially-scaled} for the kernel function, we can write 
\begin{align}
\mathcal{K}_{m}(\eta_{T} + n L/\lambdabar_{T}, s)
= 
\tilde{\mathcal{K}}^{(T)}_{m}(\eta_{T}, s) e^{-2\eta_{T}} e^{-2 n \frac{L}{\lambdabar_{T}}} 
\end{align}
where we defined $\tilde{\mathcal{K}}^{(T)}_{m}(\eta_{T}, s)\equiv\tilde{\mathcal{K}}^{(T)}_{m}(\eta_{T} + n L/\lambdabar_{T}, s)$, i.e. the scaled kernel function in terms of the shifted variable $\eta_{T}$.
The previous expression explicitly features an exponentially decaying factor in both $\eta_{T}$ as well as the summation index $n$. 
The latter occurs with a rate of $\mu_{T} = 2L/\lambdabar_{T}$, scaling linearly with the temperature $T$. 
In this form, we can thus apply the MDL Gaussian summation for the sum over $n$ according to \cref{eq:MDL-gaussian-summation}, while we utilize once more a Gauss-Laguerre quadrature rule for the integral over $\eta_{T}$. 
As previously discussed, the MDL summation is particularly effective for the low-temperature regime where a direct summation converges very slowly. 
In practice, we switch to the MDL scheme for the sum over $n$ if $L < \lambdabar_{T}$, and again switch to the MDL scheme for the sum over $m$ if $s > 1$.  In all other cases, direct summation is employed.

\end{appendices}

%\bibliography{casimir-polder-interaction-hollow-core-fiber}
%\bibliographystyle{prstytitlenew}

\end{document}